\documentstyle[psfig,amssymb,amsfonts,12pt,preprint,tighten,aps]{revtex}

\begin{document}
\input psfig.sty
\title{Update on neutrino mixing in the early Universe}
\author{P. Di Bari}
\maketitle
\begin{center}
{\em Deutsches Elektronen-Synchrotron DESY, 22603 Hamburg, Germany} \\
(dibari@mail.desy.de)
\end{center}
\date{}
\maketitle
\begin{abstract}
{From the current cosmological observations of CMB and nuclear abundances 
we show, with an analytic procedure, 
that the total effective number of extra neutrino species  
$\Delta N_{\nu}^{\rm tot}< 0.3$. We also describe the possible 
signatures of non standard effects that could be revealed in future CMB observations. 
This cosmological information is then applied to neutrino mixing models. 
Taking into account the recent results from the SNO and SuperKamiokande 
experiments, disfavouring pure active to sterile neutrino oscillations,
we show that all 4 neutrino mixing models, both of 2+2 and 3+1 type, 
lead to a full thermalization of the sterile neutrino flavor. 
Moreover such a sterile neutrino production excludes the possibility of an electron 
neutrino asymmetry generation and we conclude that $\Delta N_{\nu}^{\rm tot}\simeq 1$, 
in disagreement with the cosmological bound. 
This result is valid under the assumption that the initial 
neutrino asymmetries are small. We suggest the existence of a second sterile
neutrino flavor, with mixing properties such to generate a large 
electron neutrino asymmetry, as a possible way out. }
\end{abstract}

\newpage

\section{Introduction}

Neutrino mixing is the simplest explanation of the data from
the atmospheric \cite{atm} and solar neutrino experiments \cite{solar,SNO},
while alternative mechanisms are becoming more and more unlikely \cite{noexotic}. 
The results from the LSND experiment can be also 
explained by neutrino mixing \cite{LSND}.

In this work we investigate the possible cosmological
effects of neutrino mixing. The theory of big bang nucleosynthesis
has been proposed since long time  as a probe for particle physics 
\cite{Shvartsman,Schramm}, but the systematic uncertainties in the 
measurements of nuclear abundances represent an obstacle for 
improvements. In the last two years
different experiments confirmed the existence of acoustic peaks in the 
power spectrum of CMB temperature anisotropies \cite{CMB}, 
from which it has been possible to measure, with improved precision, many
different cosmological parameters \cite{CMB2,BOOM,DASI,MAX}. The precision 
of these measurements will be further improved by the new satellite
experiments: the MAP satellite, already launched and on the way to its
final orbit about the  L2 Lagrangian point \cite{MAP} and the PLANCK satellite whose launch
is scheduled in the year 2007 \cite{PLANCK}. These new observations
represent a way to integrate the nuclear abundances observations 
partly overcoming the obstacle of systematic uncertainties and thus offering
new opportunities to detect or constraint new physics in the 
early Universe, in particular the effects of neutrino mixing.

In the section II we describe a simple new analytic and graphical procedure
to confront a large class of possible non standard effects 
with the cosmological observations. We consider both the present 
situation, finding that the total effective number
of neutrino species $\Delta N_{\nu}^{\rm tot}<0.3$, 
but we also point out which results from future observations
could be interpreted as signatures of non standard effects. 
In sections III-IV-IV we examine the specific predictions of those neutrino mixing models
that have been proposed to explain the current data, 
including or excluding the LSND experiment and, with the new procedure, we
confront them with the cosmological observations.
We explain why the early Universe encounters difficulties in 
detecting effects from three ordinary neutrino mixing models (section III), 
while it is emphasized the unique capability of the early Universe to probe a mixing
with a new sterile neutrino sector even for very small mixing angles, 
otherwise out of the reach of Earth experiments (section IV). In the case of 
four neutrino mixing models (section V) we study the cosmological predictions
using the results obtained in the simple active-sterile neutrino mixing and neglecting
the possible presence of phases. We find the remarkable results that 
the new data from the SNO and SK experiments favour those four neutrino mixing 
models, both of `2+2' and `3+1 type', 
that lead to a final $\Delta N_{\nu}^{\rm tot}\simeq 1$. 
In section VI we show how an additional sterile neutrino flavor could 
solve the disagreement with the cosmological bound if its mixing
is able to generate a large electron neutrino asymmetry generation that 
produces a negative contribution to $\Delta N_{\nu}^{\rm tot}$.
 In section VII we conclude summarizing which are the possible signatures
of neutrino mixing models that should be searched in future observation.

\section{Cosmological observations}

\subsection{Current constraints}

The recent observations of CMB anisotropies \cite{BOOM,DASI,MAX} provide a 
useful consistency test for the other cosmological observations. The interpretation 
of data depends on theoretical assumptions. Therefore, it is important that the simplest  
model used to fit the data, that makes use of 7 independent parameters, gives results 
that are consistent with the other observations. A combined analysis 
of the experiments allowing also for the presence of tensor fluctuations 
and a hot dark matter component by increasing the number of parameters  to 11 \cite{WTZ}, 
does not show hints of the presence of such components and is remarkable that,
when information from galaxy clustering is added, an upper bound of $4.4$ eV 
on the sum of neutrino masses is found. 
Although these analysis support a cosmological consistency, 
one has to be aware that maybe we are excluding important possible effects that are still compatible with the data or that maybe some cosmological observations are affected by
systematic uncertainties and are misleading us to wrong conclusions 
and to exclude important pieces of the picture.
We will take the attitude to consider the simplest results 
as reasonable but at the same time we will check whether these assumptions 
are compatible with the neutrino mixing models that we will examine. 

In this section, we attempt to quantify the possibility that some non standard 
effects of BBN arise from neutrino mixing models.
In order to test neutrino mixing models, 
CMB anisotropies are particularly important because they
provide a measurement of the baryon content. This information has an 
important role in constraining the presence of new physics when
is taken into account in models of Big Bang Nucleosynthesys (BBN). 
The BOOMERanG and DASI collaborations find an identical value \cite{BOOM,DASI} 
\footnote{We indicate 1$\sigma$ errors for all quantities unless differently
explicitly indicated. More precisely the DASI experiment quotes
at 68\% c.l $(\Omega_b\,h^2)^{CMB}=0.022^{+0.035}_{-0.033}$.}:
\begin{equation}\label{Omega} 
(\Omega_b\,h^2)^{CMB}=0.022^{+0.004}_{-0.003} 
\end{equation} 
while the MAXIMA collaboration finds $(\Omega_b\,h^2)^{CMB}=0.0325^{+0.0125}_{-0.0125}$ at $95\%$ c.l. \cite{MAX}.  A combined analysis has been performed in \cite{WTZ} 
in which both hot dark matter and tensor fluctuations are allowed 
and it gives for the CMB alone at $95\%$ c.l. $(\Omega_b\,h^2)^{CMB}= 0.02^{+0.06}_{-0.01}$. 
If information from the IRAS PSCz survey on galaxy clustering is used then they find, at
$68\%$ c.l., $(\Omega_b\,h^2)^{CMB}=0.020^{+0.003}_{-0.003}$ \cite{private}.  This result practically coincides with (\ref{Omega}), even though different assumptions have been used. 
Therefore this seems quite a stable and reasonable value to be used for our
analysis.

The standard BBN model (SBBN) assumes the particle physics
content of the standard model of particle physics (in particular zero masses
and no mixing for neutrinos). Moreover it assumes that
the neutrino distributions are described by the Fermi-Dirac ones with zero chemical
potentials and with a temperature $T_{\nu}\propto R^{-1}$ \cite{Wagoner}.
The predicted primordial nuclear abundances are functions of the only parameter 
$\eta$, the baryon to photon ratio, related to $\Omega_b\,h^2$ by the simple relation
$\eta_{10}\equiv 10^{10}\,\eta \simeq 273.6\,\Omega_b\,h^2$. 
The value (\ref{Omega}) for $(\Omega_b\,h^2)^{CMB}$ corresponds to have
\footnote{From this moment we will always 
show values of $\eta$ in units of $10^{-10}$, 
omitting the subscript `$10$' to simplify the notation.}: 
\begin{equation}\label{eta}
\eta^{CMB}=6.0^{+1.1}_{-0.8}
\end{equation}
These predictions have to be compared with the measured values. 
A first group finds `high' values for the primordial Helium 
abundance \cite{Izotov}:
\begin{equation}\label{high}
Y_p^{\rm exp}=0.244\pm 0.002,
\end{equation} 
while a second group finds `low' values \cite{Olive}:
\begin{equation}\label{low}
Y_p^{\rm exp}=0.234\pm 0.003.
\end{equation} 
At the moment there is a tendency to admit that there are systematic uncertainties 
in this kind of measurements and to unify the two ranges of values. 
We however prefer to continue to distinguish the two different measurements.

The primordial Deuterium abundance is measured in Quasars absorption systems at high
redshift. This kind of measurements gives the result \cite{O'Meara}: 
\begin{equation}\label{D}
(D/H)^{\rm exp}=(3.0\pm 0.4)\times 10^{-5}
\end{equation} 
We will not consider
measurements of the primordial Lithium abundance since it is not fully understood 
whether we are really able to estimate how stellar processes
could have modified it to the present. A test for the SBBN means to check whether
the following conditions are fullfilled:
\begin{equation}
Y_p^{SBBN}(\eta^{CMB})  = Y_p^{\rm exp} 
\end{equation}
\begin{equation}
(D/H)^{SBBN}(\eta^{CMB})  = (D/H)^{\rm exp}
\end{equation}
The functions $Y_p^{SBBN}(\eta)$ and $(D/H)^{SBBN}(\eta)$
do not have exact analytical expression but fits around
$\eta=5$ give the results \cite{Lopez,Walker,Sarkar96}
\footnote{We are considering the neutrino heating from $e^{+}-e^{-}$ annihilations
as a non standard effect (see discussion later on) and thus we are subtracting
this contribution ($\Delta Y_p\simeq 1.4\times 10^{-4}$ \cite{Dolgov}) 
from the result found in \cite{Lopez}: 
$Y_p^{SBBN}(\eta=5,\tau_n=887\,{\rm sec})=0.2467$, where $\tau_n$ 
is the neutron life-time.}:
\begin{equation}
 Y_p^{SBBN}(\eta)   \simeq  0.2466+0.01\,\ln(\eta/5)  \label{DSBBN}
\end{equation}
\begin{equation}
(D/H)^{SBBN}(\eta)  \simeq  3.6\cdot 10^{-5}\,(\eta/5)^{-1.6} 
\end{equation}
Using the CMB value (\ref{eta}) for $\eta$, one finds that the SBBN predicts:
\begin{equation}
Y_p^{SBBN}(\eta^{CMB})  = 0.2484^{+0.0017}_{-0.0014} 
\end{equation}
\begin{equation} 
(D/H)^{SBBN}(\eta^{CMB}) =  (2.7\pm 0.7)\times 10^{-5}
\end{equation}
If we compare these values with the experimental measurements we see that the SBBN 
is in agreement with the observations if high values of $Y_p$ are used, otherwise for
low values of $Y_p$ there is a $4\,\sigma$ discrepancy. Such a comparison of SBBN 
with the observations can be also done saying that SBBN predicts, from 
the current $Y_p^{\rm exp}$ and $(D/H)^{\rm exp}$,
the following values for $\eta$ (3\,$\sigma$):
\begin{eqnarray}
\eta^{\rm SBBN}_{\rm high Y_p} & = &  3.8^{+3.2}_{-1.8}  \label{etahY} \\
\eta^{\rm SBBN}_{\rm low Y_p}  & = &  1.4^{+2.1}_{-0.8} \\
\eta^{SBBN}_{(D/H)}            & = &  5.6^{+2.1}_{-1.1} \label{etaD}
\end{eqnarray}
and comparing them with $\eta^{CMB}$ the same previous conclusions follow.
 
We want now to quantify the possibility that BBN is non standard 
and of course, in doing this, we will be particularly interested in those 
non standard BBN models that can result from neutrino mixing. 
In this case the possible non standard effects are of two kinds
and quite well known. The {\em first effect} is the possibility that the 
number of energy density degrees of freedom $g_{\rho}\equiv (30/\pi^2)\,(\rho/T^4)$ 
differs from its SBBN value $g_{\rho}^{SBBN}=(22/4)+(21/4)(T_{\nu}/T)^{4}$ 
before or during the BBN period. In this 
way the expansion rate and the standard BBN predictions for the 
primordial nuclear abundances would be modified \cite{Shvartsman}. 
The change of $g_{\rho}$ can be expressed in terms of the
(effective) extra number of neutrino species \cite{Schramm} $\Delta N_{\nu}^{\rho}$:
\begin{equation}
g_{\rho}=g_{\rho}^{SBBN}+{7\over 4}\,\Delta N_{\nu}^{\rho}\,\left({T_{\nu}\over T}\right)^4
\end{equation}
From the definition of $g_{\rho}$ it follows that $\Delta N^{\rho}_{\nu}$ 
is related to the neutrino energy densities by the following simple expression:
\begin{equation}
\Delta N^{\rho}_{\nu}=\sum_{X}\,{\rho_X+\rho_{\bar{X}}\over\rho_0}-3
\end{equation}
where $\rho_0=(7\,\pi^2/ 120)\,T_{\nu}^4$ and the `X'-particles   include 
the three ordinary neutrinos plus possible new species (we
will be interested to possible new sterile neutrino flavors).
Again we can make use of linear fits that account for the contribution
of a non zero $\Delta N_{\nu}^{\rho}$ in the BBN predictions 
for the primordial nuclear abundances 
\footnote{The number 0.0137 can be inferred from 
the expansion given in \cite{Lopez} for $\eta\simeq 6$. While
the dependence of $(D/H)$ on $\Delta N^{\rho}_{\nu}$ can be easily
calculated considering that this abundance stays constant for
$\eta/\sqrt{g_{\rho}}={\rm const}$ \cite{Lisi}.}:
\begin{equation}
 Y_p^{BBN}(\eta,\Delta N^{\rho}_{\nu})   \simeq 
Y_p^{SBBN}(\eta) + 0.0137 \,\Delta N_{\nu}^{\rho} 
\end{equation}
\begin{equation}
(D/H)^{BBN}(\eta,\Delta N^{\rho}_{\nu})  \simeq 
(D/H)^{SBBN}(\eta)\,(1+0.135\,\Delta N_{\nu}^{\rho})^{0.8}
\end{equation}
A {\em second class of deviations from SBBN} are those related to distortions of
electron neutrino and anti-neutrino SBBN distributions, given by
the Fermi-Dirac distribution with zero chemical potential 
(the same for neutrinos and anti-neutrinos). In general deviations
cannot be described in terms of a finite number set of parameters but 
by an infinite number of parameters (the occupation numbers for
each quantum state with a given momentum). 
However one can first calculate the change in $Y_p$ caused by these
deviations and then normalize this change by introducing the quantity:
\begin{equation}
\Delta N_{\nu}^{f_{\nu_e}}\equiv 
{[Y_p(\eta,\Delta N_{\nu}^{\rho},\delta f_{\nu_e,\bar{\nu}_e})-Y_p(\eta,\Delta N^{\rho}_{\nu})]\over 0.0137}
\end{equation}
In this way one weighs the effect of distortions in terms of
the presence of extra number neutrino species. 
A specific model of non standard BBN should be able to specify the deviations 
$\delta f_{\nu_e,\bar{\nu}_e}(p,t)$ at each momentum and during all the period of BBN. 
However it has to be remarked that once that the neutron to proton ratio has 
frozen, the electron neutrino ditributions do not have any more a direct role
in the nuclear reactions. Thus everything will depend only on
the frozen value of $n/p$ and still on $\Delta N_{\nu}^{\rho}$. 
This means that the Deuterium abundance will depend only indirectly on
the electron neutrino distortions through the quantity 
$\Delta N_{\nu}^{f_{\nu_e}}$. Actually  such a dependence is very weak
and we will neglect it. Of course different $\delta f_{\nu_e,\bar{\nu}_e}$ can
produce the same $\Delta N_{\nu}^{f_{\nu_e}}$ and this degeneracy represents a 
lost of information
\footnote{The only way to have more information on the $\delta f_{\nu_e,\bar{\nu}_e}$ 
would be to detect the electron relic neutrino distributions from which one 
could infer their values during BBN. Unfortunately relic neutrinos detection  
appears at the moment beyond the current observations but there are some
interesting developments from the study of UHE$\nu$ scattering on relic neutrinos 
and producing  Z bursts \cite{Andreas}.}.
 The predictions of such non  standard BBN models can again be compared 
with the experimental observations:
\begin{equation}
Y_p^{BBN}(\eta^{CMB},\Delta N^{\rho}_{\nu},\delta f_{\nu_e,\bar{\nu}_e}) 
=  Y_p^{\rm exp} 
\end{equation}
\begin{equation} 		    	
(D/H)^{BBN}(\eta^{CMB},\Delta N^{\rho}_{\nu})  =  (D/H)^{\rm exp}
\end{equation}
The $Y_p$ measurement puts constraint on the quantity
\footnote{Note there could be other kinds of non standard effects 
non considered here, like those ones
associated with the possibility that during the BBN epoch there were
baryon inhomogeneities on the scale of neutron diffusion length 
(see \cite{Hannu} and references therein). 
The quantity $\Delta N_{\nu}^{\rm tot}-\Delta N_{\nu}^{\rho}$
would assume a more general interpretation and depend also on other
non standard parameters like the size of inhomogenenities and thus
should be more generally indicated for example with 
$\Delta N_{\nu}^{f_{\nu_e}+{\rm inh}}$.   In this paper we are interested
to focus on non standard effects from neutrino mixing models and thus
we completely neglect the possibility for these kinds of 
inhomogeneities but it is interesting that
this procedure could be employed also in a different context.}: 
\begin{equation}
\Delta N_{\nu}^{\rm tot}=\Delta N_{\nu}^{\rho}+\Delta N_{\nu}^{f_{\nu_e}}\equiv \Delta Y_p^{BBN}/0.0137
\end{equation} 
At $3\,\sigma$, assuming high values for $Y_p^{\rm exp}$, we find
:
\begin{equation}\label{Ntoth}
\Delta N_{\nu}^{\rm tot}=-0.3^{+0.6}_{-0.6},
\end{equation}
while assuming low values we find:
\begin{equation}\label{Ntotl}
\Delta N_{\nu}^{\rm tot}=-1.05\pm 0.75
\end{equation}
The Deuterium abundance provides a complementary information 
on $\Delta N^{\rho}_{\nu}$ and the comparison between the prediction
and the observed value gives, conservatively at a 3$\sigma$ level, 
an upper bound on $\Delta N^{\rho}_{\nu}$:
\begin{equation}\label{Nrho}
(\Delta N^{\rho}_{\nu})^{BBN}\lesssim 13
\end{equation}
while a lower bound is still not obtained with the current precision
of measurements. The constraints (\ref{Ntoth}), (\ref{Ntotl}) and (\ref{Nrho}) 
are shown in {\bf figure 1}, in a plot $\Delta N_{\nu}^{\rm tot}-\Delta N_{\nu}^{\rho}$. 

\subsection{Future observations and possible signatures}

It is interesting to see how one can expect that these constraints will improve
from future CMB measurements of $\eta$. The Planck experiment should be able
to measure $\eta$ with a precision at the level of 1\% or less \cite{Bond}. 
In this way the uncertainties on the theoretical predictions 
of the nuclear abundances, $Y_p^{BBN}$ and $(D/H)^{BBN}$,  
will become negligible compared to the errors on the experimental values.

Assuming that the future measured value $\eta^{CMB}$ will correspond
to the current central value of $(D/H)$ in SBBN, $\eta= 5.6$ (see Eq.(\ref{etaD}) ), 
then the current Deuterium observations will
constraint $\Delta N_{\nu}^{\rho}$ to be $\lesssim 4.0$ ($3\,\sigma$)
(the horizontal thick dashed line in figure 1) while still one does not get
a lower bound
\footnote{One finds $\Delta N_{\nu}^{\rho}\gtrsim -4$
that is not particularly meaningful. At $2\,\sigma$ one gets 
$\Delta N_{\nu}^{\rho}\gtrsim -2$, that implies 
the presence of at least one standard neutrino species. Note that
a model in which a ${\rm MeV}\,\tau$ neutrino was decaying prior the onset 
of BBN was proposed to solve the BBN crisis from low $Y_{p}^{\rm exp}$ values 
\cite{Hata}. In such a case one can get negative values of $\Delta N_{\nu}^{\rho}$
as low as $-1$. Nowaday such a model is very disfavored by the neutrino oscillations
experiments but nevertheless it gives an example of why it is not 
meaningless to put a negative lower bound on $\Delta N_{\nu}^{\rho}$,
that moreover can be considered a sort of consistency check of the 
basic BBN assumptions.}. The same 
exercise can be performed with $Y_p$ 
to see how the constraints on $\Delta N_{\nu}^{\rm tot}$ would improve and
the result is shown with vertical dashed lines in {\bf figure 1}.
This time the improvement is slight because
the $Y_p$ abundance is much less sensitive to $\eta$ than $(D/H)$  
and the error on $\Delta N_{\nu}^{\rm tot}$ is dominated by the error
on $Y_p^{\rm exp}$. The two constraints together, from $(D/H)$ and $Y_p$, 
give the gray region in {\bf figure 1} with thick dashed line contours. 

Differently it could happen that future CMB observations
will indicate $\eta^{CMB}>\eta^{SBBN}_{\rm high Y_p^{\rm exp}}$  
(see Eq.  (\ref{etahY})). In {\bf figure 1} it is shown, in light gray, 
the allowed cosmological
region (at $3\,\sigma$) in the plane $\Delta N_{\nu}^{\rm tot}-\Delta N_{\nu}^{\rho}$
for $\eta^{CMB}= 7.0$ ($1\%$ error) for a `low+high' joint range of $Y_p$ values.
The SBBN would be ruled out at $3\,\sigma$ and negative 
$\Delta N_{\nu}^{\rm tot}< \Delta N_{\nu}^{\rho}$ would be required.
 Therefore, in future, a $1\%$ error measurement $\eta^{CMB}\gtrsim 7$ 
(or $(\Omega_b\,h^2)^{CMB}\gtrsim 0.0256$), will represent the opportunity 
to have a significant signature of non standard BBN effects with
the current nuclear abundances observations. 
On the other hand the current allowed  $3\,\sigma$ range of values 
of $\eta^{CMB}$, $3.6-9.9$ (see Eq. (\ref{eta})), excludes already now 
the possibility that a future $1\%$ error measurement of $\eta^{CMB}$,
with current $Y_p$ measurements, 
can give indications for positive values of $\Delta N_{\nu}^{f_{\nu_e}}$, 
that means $\Delta N_{\nu}^{\rm tot}<\Delta N_{\nu}^{\rho}$, because it would 
require $\eta^{CMB}\lesssim 2$ using high $Y_p$ values (see Eq. \ref{etahY})
and even lower values of $\eta^{CMB}$ using low $Y_p$ values.

From the Eq. (\ref{etaD}), 
one can see that from a $1\%$ error measurement $\eta^{CMB}\gtrsim 7.7$ 
the Deuterium abundance would require also $\Delta N_{\nu}^{\rho}>0$
(other than negative $\Delta N_{\nu}^{f_{\nu_e}}$).
On the converse, for $\eta^{CMB}\lesssim 4.5$, negative values of $\Delta N_{\nu}^{\rho}$
would be required. 
  
Another important improvement, from future observations of CMB anisotropies, 
will be the direct measurement of $\Delta N^{\rho}_{\nu}$.
The presence of an extra radiative component changes the
CMB spectrum, in particular leading to the enhancement of
the height of first acoustic peak.  At the
present, a completely independent measurement of $\Delta N^{\rho}_{\nu}$ 
from CMB anisotropies gives a very loose constraint 
$(\Delta N_{\nu}^{\rho})^{CMB}< 19$ (at $95\%$ c.l., CMB alone) \cite{Hannestad}. 
However future MAP and Planck satellite experiments  should reach a precision of
$10^{-3}-10^{-1}$ according to whether the 
information on the other cosmological parameters from
other observations will be used or not and whether the CMB polarization will be measured
or not \cite{Lopez2}. 
In {\bf figure 1} we indicated with horizontal thin dashed lines
a realistic future constraint $|(\Delta N_{\nu}^{\rho})^{CMB}|< 0.1$. 
It has to be said that the CMB observations will measure
$\Delta N_{\nu}^{\rho}$ around the time of recombination and thus it could be
in principle different from the value of $\Delta N_{\nu}^{\rho}$ during 
the earlier period of BBN if some intervening effect modified it.
For example $(\Delta N_{\nu}^{\rho})^{\rm CMB}$ can be higher of 
$(\Delta N^{\rho}_{\nu})^{\rm BBN}$ in the case of massive neutrino decays. 
In this case a comparison between the two quantities will 
test the `relativity parameter' $\alpha\propto m_{\nu}^2\,\tau$
\cite{Hannestad2}. 
 
Their comparison could give also another result: 
$(\Delta N_{\nu}^{\rho})^{\rm CMB}< (\Delta N_{\nu}^{\rho})^{\rm BBN}$.
This is possible only if one can say that $(\Delta N_{\nu}^{\rho})^{\rm BBN}>0$.
If one looks at the expressions (\ref{DSBBN}) and (\ref{etaD}),
such a conclusion is possible if future $1\%$ error observations will
give $\eta^{CMB}\gtrsim 7.7$. For example from $\eta^{CMB}\gtrsim 7.8$
one can deduce $(\Delta N_{\nu}^{\rho})^{BBN}\gtrsim 0.2$, while it can happen
at the same time that CMB constraints $\Delta N_{\nu}^{CMB}\lesssim 0.1$.
This paradoxical situation could occur if
$\Delta N_{\nu}^{\rm tot}$ is inhomogenous. We neglected a dependence
of $(D/H)^{BBN}$ on $\Delta N_{\nu}^{\rm tot}$, that means on $Y_p$
or, equivalently, on the frozen value of neutron to proton ratio.
 This because the $Y_p$ observations suggest that $Y_p$ cannot differ
from the SBBN value so much to modify $(D/H)$ in a sensible way, while
the value of $\Delta N_{\nu}^{\rho}$ is much less constrained and it
can considerably alter the value of $(D/H)$.
However the observations measure $Y_p$ only within about $100\,{\rm Mpc}$
around us, while the $(D/H)$ abundance is measured in the quasars absorption systems
at much larger distances. Thus it cannot be excluded that $Y_p$ can be `there' 
much larger than what we observe around us \cite{Pagel}. The amplitude of 
CMB anisotropies exclude the possibility that this spatial variation can be
due to a inhomogenous $\Delta N_{\nu}^{\rho}$,
and thus it can be due only to a inhomogeneous $\Delta N^{f_{\nu_e}}_{\nu}$
that should be `there' much larger (and positive) than around us.
In this case the $(D/H)$ nuclear abundance can
be higher than what predicted by SBBN and compatible with $\eta^{CMB}>\eta^{SBBN}$.
Such a possibility should be however accompanied by the observation of 
dispersion in the $(D/H)$ measurements in the range of values 
$(1.8-3.6)\times 10^{-5}$. 
Note that at the moment values of $\eta^{CMB}\gtrsim 7.7$ are already 
excluded at $1.5\,\sigma$ and thus a small improvement in the measurement
precision of $\eta^{CMB}$ should be able to disfavour (or reveal !) 
such a situation. However only constraining the dispersion in the 
values of measured $(D/H)$ can put more general limits
on the presence of inhomogeneities in $\Delta N_{\nu}^{f_{\nu_e}}$.

Another important possibility is whether future observations will
indicate $\Delta N_{\nu}^{\rho}>0.3$, because then, in order not to violate 
the bound $\Delta N_{\nu}^{\rm tot}<0.3$,
one can conclude that there is a negative contribution $\Delta N_{\nu}^{f_{\nu_e}}$.

\subsection{Two special cases of non standard effects}

The SBBN corresponds, in the plane $\Delta N_{\nu}^{\rm tot}-\Delta N_{\nu}^{\rho}$,
to the origin. One can consider the correction to the approximation of 
full neutrino decoupling at the time of electron-positron annihilations
(implying $T_{\nu}\propto R^{-1}$). It has been shown that 
such a correction yields $\Delta N_{\nu}^{\rm tot}\simeq 0.012$ 
and $\Delta N_{\nu}^{\rho}\simeq 0.034$ \cite{Dolgov}. 
Thus the predictions of nuclear abundances within the standard model of particle 
physics do not exactly coincide with those of SBBN.
In the optimistic case that future CMB
observations will be able to detect  $\Delta N^{\rho}_{\nu}$ as small as 
$0.01$, then the small effect of neutrino heating should be distinguished \cite{Lopez2}
and this would represent an important confirmation of the early Universe standard scenario 
below $\sim 10{\rm MeV}$. 

 A particular subclass of non standard BBN models, of the type
considered here, is that one in which neutrinos have still
thermal distributions but with non zero chemical potentials 
(fulfilling the chemical equilibrium condition 
$\mu_{\bar{\nu}_{\alpha}}=-\mu_{\nu_{\alpha}}$).
In this particular case one has the following correspondence:
\begin{equation}\label{Nxi}
\Delta N^{\rho}_{\nu}=\sum_{\alpha}
\left[{30\over 7}\left({\xi_{\alpha}\over\pi} \right)^2+
{15\over 7}\left(\xi_{\alpha}\over\pi \right)^4\right],
\hspace{10mm}
\Delta N_{\nu}^{f_{\nu_e}}\simeq -16\,\xi_e
\end{equation}
with $\xi_{\alpha}\equiv\mu_{\alpha}/T$ 
\footnote{The second relation is a good approximation for 
$|\xi_e|\ll 1$ and $\Delta N_{\nu}^{\rho}\ll 20$.}.
This kind of models have been largely studied in the literature 
since early times \cite{Wagoner,Reeves}. Constraints 
from the most recent cosmological observations have been also recently
obtained in \cite{Kneller}. Their procedure put 
constraints on $\Delta N_{\nu}^{\rho}$, assuming that is equal in
BBN and CMB epochs, and on $\eta$ in a statistical combined
way and taking into account a slight dependence of $\eta^{CMB}$
on $\Delta N_{\nu}^{\rho}$. This allows to get more restrictive constraints 
but in a more specific context and at the expenses of physical insight. 
In our procedure we get more conservative constraints becasue of the poorest
statistical procedure. On the other hand we gain more physical insight from an analytical
procedure valid in a more general framework in which we distinguish $\Delta N_{\nu}^{\rho}$
in BBN and CMB, the role of $\Delta N_{\nu}^{f_{\nu_e}}$ is emphasized
as we need for our purposes, and we find a bound on $\Delta N_{\nu}^{\rm tot}$ 
missing in \cite{Kneller}. All these features are important for our 
following considerations. 

\section{Three ordinary neutrino mixing}

With the exclusion of the LSND experiment,
usually justified with the argument that
it is the only experiment not yet confirmed by a second one, 
a three ordinary neutrino mixing can explain the current data 
from solar and atmospheric neutrino experiments. The three ordinary neutrino flavors
eigenstates, $|\nu_{\alpha}\rangle$ ($\alpha=e,\mu,\tau$), are connected to
three mass eigenstates $|\nu_i\rangle$ with definite masses $m_i$ ($i=1,2,3$), 
by a $3\times 3$ neutrino mixing matrix $U$:
\begin{equation}
|\nu_{\alpha}\rangle = \sum_{i=1}^{3} U^{\star}_{\alpha i}\,|\nu_{i}\rangle
\end{equation}
The atmospheric neutrino experiments are then explained by the
mixing of $|\nu_2\rangle$ and $|\nu_3\rangle$ mass eigenstates with 
$|m^2_3-m^2_2|=\delta m^2_{\rm atm}\simeq 2.5\times 10^{-3}\,{\rm eV}^2$
with a large mixing angle $\sin^2 2\theta_{23}\gtrsim 0.88$ \cite{atm} 
and with a negligible $|U_{e3}|\ll 1$ 
(as required by the CHOOZ experiment \cite{CHOOZ})
that implies a small mixing angle $\theta_{13}$.
In this way the $\nu_{\mu}$'s are converted almost only to $\nu_{\tau}$'s.
The solar neutrinos are explained by the mixing of $\nu_1$ and $\nu_2$ eigenstates
with $|m^2_{2}-m^2_{1}|=\delta m^2_{\bigodot}\ll 10^{-3}\,{\rm eV}^2$. 
With the new data from the SNO experiment, large mixing angle   
solutions ($\sin^2 2\theta_{12}\simeq 1$) are also favoured \cite{Fogli,Bahcall}.
In this way the favoured three neutrino mixing models are those 
close to the the bimaximal mixing scenario \cite{bimaximal}.
A three ordinary neutrino mixing does not have relevant effects on the 
cosmological picture and in particular on the quantities $\Delta N_{\nu}^{\rho}$
and $\Delta N_{\nu}^{f_{\nu_e}}$. It has been noted that a mixing of electron neutrinos
with muon/tauon neutrinos during the period of freeze out of the neutron to proton ratio
would exchange the abundances of the two types that are slightly different due to 
the different effect of neutrino heating \cite{Langacker}. In this way 
the effect of neutrino heating would change. 
However the neutrino heating effect is small like also the difference
between the muon/tauon and the electron neutrino populations. Thus 
the bimaximal mixing would represent a correction of an already 
correcting effect. Therefore, at the moment, such a mixing model seems out 
of reach of cosmological investigation 
\footnote{Cosmology is however useful to get information on the 
mass pattern, in particular structure formation and CMB observations
are currently sensitive to a few ${\rm eV}$'s masses \cite{WTZ}, while the PLANCK 
experiment will be be sensitive to a few $0.1\,{\rm eV}$'s masses \cite{Bond}.}. 

\section{Active-sterile neutrino mixing}

An explanation in terms of neutrino mixing 
of the solar and atmospheric neutrino experiments, together with
the LSND experiment, implies three different scales of mass squared differences: 
$\delta m^2_{\bigodot}\ll \delta m^2_{\rm atm}\ll \delta m^2_{\rm LSND}$.
This requires the existence of at least one new neutrino flavor \cite{Peltoniemi}
that has to be sterile in order to escape the constraint $N_{\nu}^{Z}=3.00\pm 0.06$
from the invisible decay width of the Z-boson \cite{PDG}.
A two neutrino mixing between one ordinary neutrino flavor $\nu_{\alpha}$
and one sterile neutrino flavor $\nu_s$ is the simplest case of mixing
involving new sterile neutrino flavors. 
With the new data from the atmospheric and the solar neutrino experiments, 
such  a simple scheme seems to be excluded,
as the three ordinary neutrino flavors appear to be mixed among themselves. 
However the solutions of the kinetic equations, necessary to calculate 
$\Delta N_{\nu}^{\rho}$ and $\Delta N_{\nu}^{\rm tot}$,
present many difficulties and this basic case represents
an important starting point. Moreover it can represent
a limit case for some of the possible sub-mixings  
within a realistic multiflavor mixing, as we will see in the next section. 
It is described by only two parameters, 
$\delta m^2\equiv m_{2}^{2}-m_{1}^{2}$ and $s^2\equiv \sin^2 2\theta_0$,
where $\theta_0$ is the vacuum mixing angle. The $|\nu_1\rangle$ mass eigenstate 
($|\nu_2 \rangle$) corresponds to the ordinary (sterile) 
neutrino eigenstate in the limit of no mixing. 
The straightest  cosmological effect is the sterile neutrino production
with a consequent generation of a $\Delta N_{\nu}^{\rho}$ that can be
as high as $1$ in the case of full thermalization. In doing these 
calculations one has to make some assumptions about the initial 
value of the ordinary neutrino asymmetries. We define the asymmetry of 
a lepton (baryon) particle species $X$ as:
\begin{equation}
L_{X}\,(B_{X})\equiv {N_{X}-N_{\bar{X}}\over N_{\gamma}^{\rm in}}
\end{equation}
with $N_{X}$ being the particle number per comoving volume and 
$N_{\gamma}^{\rm in}$ is the number of photons per comoving volume at
an initial temperature 
$T^{\rm in}\simeq 10\,{\rm MeV}\gg m_{\rm el}/2\simeq 0.25\,{\rm MeV}$. 
The {\em (effective) total $\alpha$-neutrino asymmetry} is defined as:
\begin{equation}
L^{(\alpha)}\equiv L_{\nu_{\alpha}}+L_{\nu_e}+L_{\nu_{\mu}}+L_{\nu_{\tau}}+Q_{\alpha}
\end{equation}
with $Q_{\mu,\tau}=-(1/2)\,B_n$ and $Q_e=L_e-(1/2)\,B_n$.
For initial values $|L^{(\alpha)}|\ll 10^{-6}\,(|\delta m^2|/{\rm eV}^2)^{1/3}$ 
(`small' neutrino asymmetries) the effects on the oscillations can be 
neglected, while for much higher values an initial neutrino asymmetry can  
modify, usually suppressing, the sterile neutrino production 
prior the onset of BBN \cite{fv,dll}. For small neutrino asymmetries,
the sterile neutrino production is given by \cite{Cline,dll}:
\begin{equation}\label{Nrhos}
N_{\nu_s}^{\rho}= 1-\exp[-g_{\pm}^{\alpha}(s^2,|\delta m^2|/{\rm eV}^2)]
\end{equation}
The function $g_{\pm}^{\alpha}(s^2,|\delta m^2|/{\rm eV}^2)$ can be written
in the form:
\begin{equation}
g_{\pm}^{\alpha}(s^2,\delta m^2/{\rm eV}^2)=
K_{\alpha}\,F_{\pm}^{\alpha}(s^2)\,s^2 \sqrt{|\delta m^2|/{\rm eV}^2}
\end{equation}
where the subscript $+$ ($-$) stands for positive (negative) $\delta m^2$
and $K_{\alpha}\simeq 657\,(898)$ for $\alpha=e\,(\mu,\tau)$. The
function $F_{\pm}^{\alpha}(s^2)$ is given by the following integral:
\begin{equation}
F_{\pm}^{\alpha}(s^2)=
\int_0^\infty
dt ~{t^2 \over s^2  + d_0^2\, t^{12} + (c \pm t^6)^2 }
\end{equation}
with $c\equiv \cos 2\theta_0$ and $d_{0}\simeq 0.008\,(0.02)$ 
for $\alpha=e\,(\mu,\tau)$. These results
have been obtained within the static approximation \cite{ftv} that
neglects the MSW effect at the resonance. In the resonant case, for negative
$\delta m^2$, this approximation holds only for very small mixing angles ($s^2\ll 10^{-4}$)
and for small neutrino asymmetries \cite{comment}. Note that 
$\Delta N_{\nu}^{\rho}=N_{\nu_s}^{\rho}+(\sum^{e,\mu,\tau}_{\beta} N_{\nu_{\beta}}^{\rho}-3)$,
where the second term takes into account the depletion of ordinary neutrinos 
that  is negligigible when the bulk of sterile neutrino production occurs 
before the neutrino chemical decoupling and in this case 
one has approximately $\Delta N_{\nu}^{\rho}\simeq N_{\nu_s}^{\rho}$. 
This is verified for $|\delta m^2|\gtrsim 10^{-4}\,{\rm eV}$
that, for values of $N_{\nu_s}\gtrsim 0.01$, corresponds to have small
mixing angles $s^2 \lesssim 10^{-2}$ in the non resonant case and 
$s^2\lesssim 10^{-4}$ in the resonant case. 
In these regimes of small mixing angles, the function $F_{\pm}^{\alpha}(s^2)$
is well approximated by its asymptotic value $F_{\pm}^{\alpha}(0)$ and one has
for $\alpha=e\,(\mu,\tau)$:
\begin{eqnarray}
g_{+}^{\alpha} (s^2\lesssim 10^{-2},1)/s^2\simeq K_{\alpha}\,F_{+}^{\alpha}(0) 
& \simeq &  1.69\,(2.33) \,10^2 \label{g+} \\
g_{-}^{\alpha}(s^2 \lesssim 10^{-4},1)/s^2\simeq K_{\alpha}\,F_{-}^{\alpha}(0) 
& \simeq &  4.28\,(2.27) \,10^{4} \label{g-}
\end{eqnarray}
In the non resonant case these analytical results agree very well
with the numerical ones found in \cite{ekt} 
\footnote{In \cite{ekt} the results are presented for $\Delta N_{\nu}^{\rm tot}$,
however for small enough mixing angles 
the contribution to $\Delta N_{\nu}^{f_{\nu_e}}$ is negligible
and a comparison is possible at least for the non resonant case.  
In the resonant case a comparison with the results of \cite{ekt},
at $s^2\geq 10^{-4}$, is not possible because these 
take into account also the negative contribution 
$\sum_{\beta}^{e,\mu,\tau} N_{\nu_{\beta}}-3$ to $\Delta N_{\nu}^{\rho}$
and are performed in a quantum kinetic formalism that accounts also for the MSW effect.}
for $s^2\geq 10^{-4}$, while in \cite{Dolgov4} it is claimed that 
$\Delta N_{\nu}^{\rho}$ is approximately three times lower. 

All these results have been obtained assuming small neutrino asymmetries.
However in the resonant case, at small mixing angles, even if one starts with
small neutrino asymmetries, a large $\alpha$-neutrino asymmetry is generated
around the critical temperature:
\begin{equation}
T_c\simeq 15.0\,(18.6){\rm MeV}\,(|\delta m^2|/{\rm eV^2})^{1/6}\,(2/y_c)^{1/3},
\end{equation}
where $y_c$ is the critical (re-scaled) momentum \cite{ftv,fv2,ropa3}.  
The growth is first driven by the neutrino collisions 
that suppress the MSW effect. When the asymmetry reaches 
a value for which the interaction length at the resonance is larger
than the resonance width, then the growth starts to be driven by the
MSW effect \cite{comment,ropa3} that can bring the $\nu_{\alpha}$-asymmetry
up to a maximum value of $0.375$. A remarkable feature is 
that the MSW effect is adiabatic for 
$s^2\gtrsim 10^{-9}\,({\rm eV}^2/|\delta m^2|)^{1/4}$ \cite{ropa3}. 
Below this value the MSW effect becomes non adiabatic
and ordinary neutrinos are not converted efficiently 
into sterile neutrinos anymore. However such a small value
of the vacuum mixing angle represents by far  
the best example of how matter effects can enhance 
the mixing in vacuum, considering that in the Sun enterior
the MSW effect occurs for $s^2\gtrsim 10^{-4}$ \cite{MS}. 

One would also like to know which is the upper limit
on the vacuum mixing angle for the neutrino asymmetry 
to be generated. This is a point that has still not been 
investigated in the literature  
but we will see, in the next section, that it
will prove to be very important for our considerations.
Fortunately it is possible to get an analytic estimation.
We will be particularly interested to values of 
$|\delta m^2|\simeq \delta m^2_{LSND}\simeq 1\,{\rm eV^2}$.
For these values one can use the Eq. (\ref{g-}) to calculate
the sterile neutrino production $N_{\nu_s}^{\rho}$.
When the sterile neutrino production is negligible
($N^{\rho}_{\nu_s}\lesssim 0.1$) the value of the critical momentum is
approximately given by the peak of Fermi-Dirac distribution: $y_c\simeq 2$ \cite{fv,dll}.
Once that the asymmetry generation has started the sterile neutrino 
production is suppressed in the collision dominated regime. 
This has the effect that the sterile neutrino production 
calculated by the Eq. (\ref{Nrhos}) is halved. 
Thus, taking into account
this effect, one can easily calculate that 
the condition $\Delta N_{\nu_s}^{\rho}< 0.1$ corresponds
to mixing angles 
$s^2< 0.52 \,(0.98) \times 10^{-5}\,\sqrt{{\rm eV}^2/|\delta m^2|}$.
When this condition is verified, together with the lower limit from the adiabaticity,
the final value of the neutrino asymmetry is very close to the impassable
limit corresponding to a situation in which,
for an initial positive (negative) value of $L^{(\alpha)}$,
all anti-neutrinos (neutrinos) are converted into anti-sterile (sterile) neutrinos  
and thus $|L_{\nu_{\alpha}}|^{\rm max}=n_{\nu_{\alpha}}\,(n_{\bar{\nu}_{\alpha}})/n_{\gamma}=3/8$
\cite{fv2}.
Therefore the maximum value is also independent on the mixing angle in this range of values.
When the sterile neutrino production becomes not negligible 
($\Delta N_{\nu}^{\rho}\gtrsim 0.1$), 
it has the effect to delay the asymmetry generation since the value of $y_c$ increases
and therefore $T_c$ decreases. When $y_c$ becomes higher than $\sim 10$ the asymmetry
generation at the critical temperature is driven by resonant neutrinos
well in the tail of the distribution. Thus it is reasonable to think that
when this happens the asymmetry generation mechanism will start to turn off.
Unfortunately it is not easy to give an analytic description of this effect.
However there is a much simpler reason for which the final value of the neutrino 
asymmetry has to decrease when the sterile neutrino production becomes not negligible.
The reason is that the final value is reached during the MSW dominated regime that
starts when the asymmetry has become large enough, during the collision dominated regime, 
that the neutrino and anti-neutrino resonances get completely separated 
and only anti-neutrino resonance can give a relevant effect, while the neutrino resonance
is by far in the tail of the distribution if $L^{(\alpha)}$ is initially positive, 
vice-versa if it is negative.
 In this way the MSW effect enhances the asymmetry to its maximum value \cite{fv2}. 
However if sterile neutrinos have been produced during the previous 
collision dominated regime, not only ordinary anti-neutrinos 
will be converted into sterile anti-neutrinos, but also the already produced 
sterile anti-neutrinos will be converted back into ordinary anti-neutrinos.
Thus the maximum value of the final neutrino asymmetry becomes 
\footnote{In \cite{dll} it has been shown that the distribution function of 
sterile neutrinos produced during the collision dominated regime, 
for $y_c\gg 1$, is given just by the equilibrium distribution times 
a coefficient $\alpha\leq 1$
in a way that  $n_{\nu_s}/n_{\nu}^{\rm eq}=
\rho_{\nu_s}/\rho_{\nu}^{\rm eq}=N_{\nu_s}^{\rho}=\alpha$.}: 
\begin{equation}\label{finL}
|L_{\nu_{\alpha}}|^{\rm max}={3\over 8}(1-N_{\nu_s}^{\rho}) 
\end{equation}
It will prove useful to assume, as upper limit on the mixing angle
for the generation of neutrino asymmetry, the value for which
$N^{\rho}_{\nu_s}>0.9$, corresponding to a final neutrino asymmetry 
{\em at least} one order of magnitude less than its maximum value 3/8. 
It is easy to calculate this value:
\begin{equation}\label{upper}
s^2 \simeq 0.5 \,(1)\times 10^{-4}\sqrt{{\rm eV}^2/ |\delta m^2|}
\end{equation}

Let us discuss now the effects of a large neutrino asymmetry.
The neutrino asymmetry generation yields 
two corrections to the $N_{\nu_s}^{\rho}$ calculated,
in the resonant case, from the Eq. (\ref{Nrhos}). 
A first correction is due to the effect, just described,
of suppression of the sterile neutrino production after the generation of the asymmetry 
and can be described by a corrective factor to the Eq. (\ref{Nrhos}) 
that can be as low as $0.5$, for $N^{\rho}_{\nu_s}\lesssim 0.1$ and 
that becomes quickly $1$ (no suppression)
for $N_{\nu_s}^{\rho} \gtrsim 0.1$ \cite{dll}. A second effect
takes into account the  sterile neutrino production in the MSW 
dominated regime that results as an additive contribution
to $N_{\nu_s}^{\rho}$ from the Eq. (\ref{Nrhos}) that
accounts only for the sterile neutrino production in the collision dominated 
regime. In the calculation of $\Delta N^{\rho}_{\nu}$
one has to take into account also the depletion of ordinary neutrinos.
For $-\delta m^2\ll 100\,{\rm eV^2}$ the contribution to the sterile
neutrino production from the MSW dominated regime occurs below the neutrino chemical decoupling and is compensated by an opposite ordinary neutrino depletion and thus there is
no contribution to $\Delta N^{\rho}_{\nu}$. 
For higher values of $-\delta m^2$ ordinary neutrino are 
re-populated by the annihilations and this second contribution
to $\Delta N^{\rho}_{\nu}$ can be as high as $0.4$ \cite{Robert}.
However values of $|\delta m^2|\gtrsim 20\,{\rm eV}^2$  
are disfavoured by structure formation + CMB considerations (see for example \cite{WTZ}). 
In any case the sum of the two contributions to the total $\Delta N_{\nu}^{\rho}$,
from the two different regimes, cannot be much higher than $1$. 
Thus the account of the neutrino asymmetry generation leads only to corrections 
to the calculation of $\Delta N_{\nu}^{\rho}$.

In the case $\alpha=\mu,\tau$ the  contribution 
$\Delta N^{\rho}_{\nu}$ to $\Delta N_{\nu}^{\rm tot}$ is the leading effect
\footnote{There is a small positive contribution to $\Delta N_{\nu}^{f_{\nu_e}}$ 
at large mixing angles and $|\delta m^2| \lesssim 10^{-4}$ due to a small depletion
of $\nu_e$ number density induced by the much higher $\nu_{\alpha}$ 
number density depletion \cite{ekt,Robert}.} and we
can approximately say that the accessible 
region in the plot $\Delta N_{\nu}^{\rm tot}-\Delta N_{\nu}^{\rho}$ 
lies along $\Delta N_{\nu}^{\rm tot}=\Delta N_{\nu}^{\rho}$
for $0\leq \Delta N_{\nu}^{\rho}\lesssim 1$ (see {\bf figure 2}).

In the case $\alpha=e$, a large $\Delta N_{\nu}^{f_{\nu_e}}$ can arise
from two different processes. A first process is the $\nu_e,\bar{\nu}_e$ 
number density depletion that this time is a direct and relevant effect occurring for 
$|\delta m^2|\lesssim 10^{-4}\,{\rm eV^2}$ and yields always a 
positive $\Delta N_{\nu}^{f_{\nu_e}}$ that can be even higher than 1
\footnote{For example in \cite{Kirillova} it is shown that, for $s^2=1$ and 
$\delta m^2\simeq 3\times 10^{-8}\,{\rm eV^2}$, the $Y_p$ production 
is $0.02$ higher than in SBBN, corresponding to $\Delta N_{\nu}^{\rm tot}\simeq +1.5$
and implying $\Delta N_{\nu}^{f_{\nu_e}}$ at least as high as $+0.5$ 
(the value of $\Delta N_{\nu}^{\rho}$ is not separately shown). 
Extrapolating to higher values of $|\delta m^2|$ it seems also quite evident
that much higher values of $\Delta N_{\nu}^{\rm tot}$ ($3,4,\dots,$?)  
and of $\Delta N^{f_{\nu_e}}_{\nu}$ ($2,3,\dots,$?) are possible. 
This is confirmed by the results of a very recent work \cite{Kirillova2} 
in which it is found that $(\Delta N_{\nu}^{\rm tot})^{\rm max}\simeq 6$, 
implying $\Delta N_{\nu}^{f_{\nu_e}}$ at least as high as $\simeq 5$.}.
The cosmological constraints are thus strongly strenghtened by 
the account of this effect \cite{constraints,ekt} and this can
be seen in the plot $\Delta N_{\nu}^{\rm tot}-\Delta N_{\nu}^{\rho}$ 
considering that the region $\Delta N_{\nu}^{\rm tot} > \Delta N_{\nu}^{\rho}$ 
lies largely outside the cosmological allowed region (see {\bf figure 2}).
The second process is the generation of a large electron neutrino asymmetry in
the resonant case and at small mixing angles. This time  
the sign is the same one of the initial $L^{(e)}$ that is observationally  
unknown and that could be predicted only within a full baryo-leptogenesys 
model and thus it can be both positive and negative. 

For negative values it is remarkable that the region 
$\Delta N_{\nu}^{\rm tot} < \Delta N_{\nu}^{\rho}$  
becomes accessible in the $\Delta N_{\nu}^{\rm tot}-\Delta N_{\nu}^{\rho}$ 
plot. In \cite{Robert} it has been calculated that
$\Delta N_{\nu}^{f_{\nu_e}}$ can be as low as $-1.4$ 
(for $-\delta m^2 \lesssim 3\,{\rm eV}^2$)
and in {\bf figure 2} one can see that  
for values $\Delta N_{\nu}^{f_{\nu_e}}\lesssim -0.3$,
there is compatibility with the region allowed by the low $Y_p^{\rm exp}$ 
values. Thus $\nu_e\leftrightarrow\nu_s$ oscillations provide 
a viable mechanism to solve the claimed SBBN crisis \cite{Shi,fv2}.
 
Another interesting possibility, 
shown in {\bf figure 2}, is that 
$\nu_e\leftrightarrow\nu_s$ oscillations would be 
also able to reconcile possible future ($1\%$ error) values of 
$\eta^{CMB}\gtrsim 7\gtrsim\eta^{SBBN}$, with the
nuclear abundances observations
\footnote{This possibility  has been proposed in \cite{ropa2}, 
when the first data from BOOMERanG-MAXIMA 
were indicating $\eta^{CMB}=9.0\pm 1.4$ \cite{BoomMax}.}.

Still another interesting effect could be the possibility to 
generate the electron neutrino asymmetry in a inhomogeneous way.
This would require the presence of small baryon number 
inhomogeneities \cite{domains}. This effect could produce
inhomogeneous nuclear abundances that could have two kind of 
indications as discussed in the previous section: indirectly
if one finds $\eta^{CMB}>\eta^{SBBN}_{D/H}$ and
$(\Delta N_{\nu}^{\rho})^{CMB}<(\Delta N_{\nu})^{BBN}$
or, directly, if one finds a dispersion in the values of $(D/H)$
measured from quasar absorption systems
\footnote{It is also interesting that inhomogeneous neutrino asymmetries,
though on much smaller scales than those necessary to produce inhomogenous nuclear 
abundances, could be responsible for the generation of galactic
magnetic fields and give rise to a detectable cosmological 
background of gravitational waves \cite{Grasso}.}.

Thus the generation of an electron neutrino asymmetry
yields many interesting cosmological effects but, 
within a two neutrino mixing scenario, it 
appears as a special possibility, considering 
that it requires $\alpha=e$, negative $\delta m^2$ 
and small mixing angles. However we saw that the generation takes
place even for tiny values of the vacuum mixing angles and because
of this, the early Universe is the most sensitive way 
to probe small mixings with new sterile neutrino flavors. 
 Moreover when considering realistic multiflavour mixing scenarios 
the conditions for the occurrence of an electron neutrino asymmetry 
generation can be more naturally satisfied.

\section{Four neutrino mixing}

Four neutrino mixing models represent the minimal way to
explain, in terms of neutrino oscillations, 
all three anomalies including the results of the
LSND experiment. These models are described by a $4\times 4$ unitary
mixing matrix $U$ connecting the 4 mass eigenstates 
$|\nu_{i}\rangle$, with definite masses $m_i$,
to the 4 flavor eigenstates $|\nu_{\alpha}\rangle$ 
($\alpha=e,\mu,\tau,s$):
\begin{equation}
|\nu_{\alpha}\rangle = \sum_{i=1}^{4} U^{\star}_{\alpha i}\,|\nu_{i}\rangle
\end{equation}

There are different possible patterns for the mass spectra but 
all of them can be distinguished in two types \cite{Okada,Bilenky,Barger0}. 
In a first type, the `3+1' models, the mass eigenvalue $m_4$ is
separated by the other three, $m_i$, by the LSND gap in a way that
$|\delta m^2_{4i}|\simeq \delta m^2_{LSND}$.
This case is a minimal modification of a three neutrino mixing model, 
since the introduction of a fourth mass eigenstate, to incorporate
the LSND results, just perturbs the mixing among the other three 
explaining solar and atmospheric neutrino results.
This means that the fourth eigenstate is almost coinciding with
the sterile neutrino flavor ($|U_{s4}|^2\simeq 1$, 
$|U_{\alpha s}|^2\ll 1$, $\alpha\neq s$). 
In a second type of models, the `2+2' models, 
the spectrum splits in two nearly degenerate 
pairs with $|\delta m^2|=\delta m^2_{\bigodot}$, $\delta m^2_{\rm atm}$, 
separated by the much larger LSND scale $\delta m^2_{LSND}$. 
In this case the neutrino mixing matrix is very different 
from the case of three neutrino mixing models. 

There is an ongoing debate on which of the two types
can better describe the experimental data \cite{Bilenky,others,Barger1,Valle,Gonzalez}. 
The new data from atmospheric neutrino experiments plus the inclusion of Tritium
$\beta$ decay data corner `3+1' models in two only allowed regions,
at $99\%$c.l., around $\delta m^2_{LSND}\sim 0.9$ and $2\,{\rm eV}^2$
\cite{Valle}. 
On the other hand the fact that 
both atmospheric \cite{SK}
\footnote{See also \cite{Rob} for a critical discussion.}
and solar neutrino \cite{SNO} data 
disfavour pure active-sterile oscillations,
suggests, for `2+2' models, that
the $\nu_e$'s (for solar) and the $\nu_{\mu}$'s (for atmospheric)
are converted into some
admixture of both active and sterile neutrinos \cite{Barger,Gonzalez}.
Thus from solar and atmospheric neutrino experiments there is no evidence 
of the existence of sterile neutrinos and the 
simplest four neutrino mixing models, predicting pure active 
to sterile neutrino oscillations, 
are disfavoured. However there is not uncompatibility among the three  
experiments when the full range of possible four neutrino mixing models
is considered. We will now study the cosmological effects of both `3+1' and `2+2'
class of models.

\subsection{3+1 models} 

The `3+1' models can be distinguished in two classes, A and B, 
such that $m_{4}\gg m_{i\neq 4}$ in A  and 
$m_{4}\ll m_{i\neq 4}$ in B. 
In the B case the three heavier mass eigenstates 
are almost degenerate with $m_i \simeq 
\sqrt{\delta m^2_{LSND}}\simeq 0.95$ or $1.4\,{\rm eV}$ according
to which of the two allowed islands is considered.
The Heidelberg-Moscow experiment on $(\beta\beta)_{0\nu}$  decay puts
restrictions on the B class \cite{Klapdor}.
 We can make use of the results seen for 
$\nu_{\alpha}\leftrightarrow\nu_s$ to get some simple estimations
on the cosmological output of the two different classes. 

{\em Class A.} One has to consider the different possible ways of 
oscillations into the sterile neutrino flavor. 
The sterile neutrino flavor almost coincides with the fourth eigenstate but 
it is also slightly present in the other three eigenstates.
The mixing between the three light eigenstates and
the fourth eigenstate is set by $\delta m^2_{4i}$ and,
since it is positive, there is no neutrino asymmetry generation. 
The mixing of the three active neutrinos with the sterile neutrino 
can be described by three different mixing angles, 
$\sin^2 2\theta_{\alpha s}\simeq 4U^{2}_{\alpha 4}$.
For $\alpha=e,\mu$ there are limits from the CDHS and BUGEY experiment
for which $\sin^2 2\theta_{\alpha s}\lesssim 10^{-1}$. For $\alpha=\tau$
we can assume the same limit. Thus from the mixing setted by
$\delta m^2_{LSND}$ and using the Eq.'s (\ref{Nrhos}) and (\ref{g+}) 
one can can see that there is a total thermalization 
($\Delta N_{\nu}^{\rho}=1$).
The LSND experiment relates the two mixing angles in a way that
$\sin^2 2\theta_{es}\times\sin^2 2\theta_{\mu s} \simeq 3\times 10^{-4}$. 
Therefore, even in the case of minimum sterile
neutrino production, when $\sin^2 2\theta_{es}=\sin^2 2\theta_{\mu s}\simeq 10^{-2}$
one has $\Delta N_{\nu}^{\rho}\simeq 0.9$, very close to a complete thermalization. 
A mixing of the three active neutrinos 
with the sterile neutrino can be also driven by $\delta m^2_{\bigodot}$
and $\delta m^2_{\rm atm}$, since the sterile neutrino is also slightly
present in the three light eigenstates. Now the sign of $\delta m^2$ can
be also negative and thus a neutrino asymmetry generation could occur in principle
 but the presence of a large sterile neutrino population,
from the mixing set by $\delta m^2_{LSND}$, will largely 
decrease the final value of the asymmetry, at least of one order of magnitude
(see the Eq. (\ref{finL})). In any case such a generation of neutrino 
asymmetry occurs for $|\delta m^2|\ll 10^{-2}\,{\rm eV}^2$ and in this case
the critical temperature would be lower than the freezing temperature 
of the neutron to proton ratio and would not affect BBN predictions in a way that
$|\Delta N_{\nu}^{f_{\nu_e}}|$ is negligible. Thus the only relevant 
cosmological effect is $\Delta N_{\nu}^{\rho}\simeq 0.9$

{\em Class B.} In this case the mixing of the three quasi-degenerate heavier
eigenstates with the fourth eigenstate has a negative 
$\delta m^2_{4i}\simeq -\delta m^2_{LSND}$.
Therefore the sterile neutrino production is 
of resonant type and from the Eq. (\ref{g-}) with $|\delta m^2|\simeq 1\,{\rm eV}^2$
and $\sin^2 2\theta \simeq 10^{-2}-10^{-1}$ one can see that again
a complete thermalization would occur with $\Delta N_{\nu}^{\rho}$ very close to $1$.
In principle an electron asymmetry generation can
also occur but the complete sterile neutrino thermalization
has the effect to suppress completely the asymmetry generation mechanism
and thus we can conclude that also in the B case 
$|\Delta N_{\nu}^{f_{\nu_e}}|\ll 1$ and therefore 
$\Delta N_{\nu}^{\rm tot}=\Delta N_{\nu}^{\rho}\simeq 1$.

\subsection{2+2 models}

These can be also distinguished in two classes, A and B. In the A (B) class  
the two lightest mass eigenstates, with masses $m_1$ and
$m_2$, explain solar (atmospheric) neutrino data while the two heavier,
with masses $m_3$ and $m_4$, explain the atmospheric (solar) 
neutrino data \cite{Bilenky,Barger0}.
Let us define simple limit cases 
in which the lightest and heaviest pair of mass eigenstates
are made only of two flavor eigenstates, that means
to consider a mixing matrix with two unmixed $2\times 2$ blocks.  
Since the atmospheric neutrino experiments constraint the probability
of $\nu_\mu\rightarrow\nu_{e}$ conversions to be very small, then one
has only four different possibilities:
\begin{enumerate} 	 
\item The $m_3,m_4$ mass eigenstates are made only of 
$\nu_{\mu}$, $\nu_\tau$ and the $m_1,m_2$ mass eigenstates
only of $\nu_{e},\nu_{s}$.
\item  The $m_3,m_4$ mass eigenstates are made only of 
$\nu_{e}$, $\nu_s$ and the $m_1,m_2$ mass eigenstates
only of $\nu_{\mu}$, $\nu_{\tau}$.
\item  The $m_3,m_4$ mass eigenstates are made only of 
$\nu_{\mu}$, $\nu_s$ and the $m_1,m_2$ mass eigenstates
only of $\nu_{e}$, $\nu_{\tau}$.
\item  The $m_3,m_4$ mass eigenstates are made only of 
$\nu_{e}$, $\nu_{\tau}$ and the $m_1,m_2$ mass eigenstates
only of $\nu_{\mu}$, $\nu_{s}$.
\end{enumerate}
Note that the 1 and 3 cases belong to the A class, while the
2 and 4 cases belong to the B class. A given neutrino flavor is
only contained in one of the two pairs of mass eigenstates,
that we call the {\em normal couple}, while it is absent
in the other one, that we call the {\em opposite couple}.
It is  is simple to calculate the cosmological output since no
neutrino asymmetry generation is possible (thus $\Delta N_{\nu}^{f_{\nu_e}}=0$)
and $\Delta N_{\nu}^{\rho}=1$ for the cases 3 and 4 and also for the cases
1 and 2 if the LMA solution is considered for the solar neutrinos
\footnote{A LMA solution for a mixing $\nu_e\leftrightarrow\nu_s$ is excluded
by the Homestake experiment but it becomes possible is some mixture of
$\nu_{\tau}$ is added to $\nu_s$ \cite{Gonzalez} or if Homestake
is disregarded \cite{Barger1}.}, otherwise $\Delta N_{\nu}^{\rho}\simeq 4\times 10^{-4}$ 
for the SMA solution ($\sin^2 2\theta_{\rm SMA}\simeq 10^{-3}$, 
$|\delta m^2|_{\rm SMA}\simeq 5\times 10^{-6}\,{\rm eV}^2$ \cite{Bahcall})  
\footnote{The  SMA solution  cannot give an 
electron neutrino asymmetry generation because 
$\delta m^2$ is positive.}.

These simple four limit cases cannot explain the experiments for two reasons.
The first reason is that in order to explain the LSND experiment the  
probability of $\nu_{\mu}\rightarrow\nu_e$ conversions 
cannot vanish and the second reason is that
the SK experiment \cite{SK}
and the SNO experiment \cite{SNO} disfavour pure 
active to sterile oscillations. 
In order to explain the LSND experiment it is necessary that a small 
mixing between the heavy and light pairs of mass eigenstates 
is introduced in a way that there is a small contamination of
$\nu_e$ and $\nu_{\mu}$ also in the respective 
{\em opposite couple} \cite{Bilenky,Barger0}. 
In order to explain the SK and SNO results one has to introduce a parameter
that allows also for $\nu_e\rightarrow\nu_{\mu,\tau}$
conversions 
\footnote{With $\nu_{\mu,\tau}$ we indicate a mixture of
$\nu_{\mu}$ and $\nu_{\tau}$. This further mixing has no relevance
in cosmology, since the $\nu_{\mu}$ and $\nu_{\tau}$ flavors cannot be
distinguished.}
in the case 1 and 2 and for $\nu_{\mu}\rightarrow\nu_{\tau}$
in the cases 2 and 4. This is usually done introducing a mixing angle between
the sterile and $\nu_{\mu,\tau}$ in a way that $(\nu_{\mu,\tau},\nu_s)\rightarrow
(\nu'_{\mu,\tau},\nu'_s)=U(\alpha)(\nu_{\mu,\tau},\nu_s)$, where $U(\alpha)$ is 
a $2\times 2$ rotation matrix \cite{Barger0}. 
In this way the 1 and 2 cases correspond to $\alpha=0$,
while the 3 and 4 cases correspond to $\alpha=\pi/2$ and for $\alpha=0 \rightarrow \pi/2$
there is a continuous transformation bringing from 1 to 3 and from 2 to 4 .

{\em Class A.} Let us first consider the transformation from 1 to 3. 
It is remarkable that  when 
the condition for adiabaticity is satisfied for very small mixing angles
$\sin^2\alpha\gtrsim 10^{-9}\,({\rm eV}^2/|\delta m^2|)^{1/4}$,
a large muon-tauon neutrino asymmetry can be generated.
This asymmetry generation can both suppress the sterile neutrino production from
$\nu_{e}\rightarrow\nu_{s}$ but also be partly transferred into an electron 
neutrino asymmetry yielding 
$\Delta N_{\nu}^{f_{\nu_e}}\simeq -0.3$ or 
$\Delta N_{\nu}^{f_{\nu_e}}\simeq 0.1$, according on
the sign of the initial total asymmetry $L^{(\mu,\tau)}$ \cite{Bell}.
However for $\sin^2 2\alpha\gtrsim 10^{-4}$ (see Eq. (\ref{upper})), the mixing 
$\nu_{\mu,\tau}\leftrightarrow\nu_s$ with $\delta m^2\simeq -\delta m^2_{LSND}$,
produces a sterile neutrino production $\Delta N_{\nu}^{\rho}\gtrsim 0.9$ 
that suppresses a large neutrino asymmetry generation and the final
$|\Delta N_{\nu}^{f_{\nu_e}}|\ll 0.1$. 
The experimental data favour, like  best fits,  
$\sin^2 \alpha \simeq 0.03-0.09$ and $\sin^2 \alpha= 0.80-0.82$ \cite{Gonzalez} 
\footnote{In the notation of \cite{Gonzalez} $\cos\alpha=c_{23}c_{24}$.}
and disfavour the possibility to have $\alpha$ 
(and also $\alpha'\equiv\pi/2-\alpha$) smaller than $10^{-4}$.
Therefore the cosmological effects are a resonant sterile neutrino production
with $\Delta N_{\nu}^{\rho}\simeq 1$ and $|\Delta N_{\nu}^{f_{\nu_e}}|\ll 1$
for $\sin^2 \alpha$ around the range $0.03-0.09$ and 
a non resonant sterile neutrino production with 
$\Delta N_{\nu}^{\rho}\simeq 1$ and $|\Delta N_{\nu}^{f_{\nu_e}}|=0$
for $\sin^2\alpha$ around  $0.80-0.82$. 
Note that the result is always $\Delta N_{\nu}^{\rho}\simeq 1$, 
even assuming a SMA solution to solve the solar neutrino problem.
In {\bf figure 3} the approximate accessible region for the class A `2+2'
models and for $\sin^2\alpha\gtrsim 10^{-9}\,({\rm eV}^2/|\delta m^2|)^{1/4}$
is shown in the $\Delta N_{\nu}^{\rm tot}-\Delta N_{\nu}^{\rho}$ plot
 with thick solid lines. The experimental results from the SNO and SK 
experiments corner it to the `point'  
$\Delta N_{\nu}^{\rho}=\Delta N_{\nu}^{\rm tot}\simeq 1$, 
represented as a small circle. 

{\em Class B.}
The other possibility is to consider a transition from the 
case 4 to the case 2 for 
$\alpha'=\pi/2-\alpha=0\rightarrow \pi/2$. Again when 
$\sin^2 2\alpha'$ becomes $\gtrsim 10^{-9}\,({\rm eV}^2/|\delta m^2|)^{1/4}$
a neutrino asymmetry generation occurs and this inhibits the sterile
neutrino production from $\nu_{\mu}\leftrightarrow\nu_s$
\footnote{Calculations of 
$|\Delta N_{\nu}^{f_{\nu_e}}|$ are missing in this case.}.
However again the SNO and SK experiments favour $\sin^2 2\alpha'\gg 10^{-4}$
and the consequent large sterile neutrino production prevents
a large neutrino asymmetry to be generated and again 
$\Delta N_{\nu}^{\rho}\simeq \Delta N_{\nu}^{\rm tot}\simeq 1$.
In {\bf figure 3} we again show a plausible accessible 
region in the $\Delta N_{\nu}^{\rm tot}-\Delta N_{\nu}^{\rho}$ plot.
Since in the case of asymmetry generation there are no calculations 
of $|\Delta N_{\nu}^{f_{\nu_e}}|$, we show the most conservative region 
(between the dashed lines) assuming that $|\Delta N_{\nu}^{f_{\nu_e}}|$ 
can take all values between zero and the maximum possible value.
This value corresponds to the case of an 
electron neutrino asymmetry generation in the limit
of two neutrino mixing $\nu_e\leftrightarrow \nu_s$,  
for $-\delta m^2\simeq 1\,{\rm eV}^2$, 
as calculated in \cite{Robert}. 

Let us try to quantify to which confidence level the results 
found in \cite{Gonzalez} constraint the possibility to have very small mixing angle 
$\sin^2\,\alpha \lesssim 10^{-4}$ or $\sin^2 2\alpha'\lesssim 10^{-4}$, 
that is equivalent to exclude the possibility of a neutrino
asymmetry generation respectively in the A class and in the B class. 
This depends on which solution one assumes for the solar neutrino data. If one assumes 
a LMA solution then the best fit is for $\sin^2 \alpha=0.80-0.82$ or
equivalently for $\sin^2 \alpha'=0.18-0.2$. Very small values  
$\sin^2 2\alpha'\lesssim 10^{-4}$ are excluded approximately at 95\% c.l.
If one assumes a SMA solution then the best fit is for $\sin^2 \alpha=0.03-0.09$.
In this case very small values $\sin^2 \alpha\lesssim 10^{-4}$ are very 
slightly disfavoured and cannot be excluded to a significant statistical confidence 
level. However, from the reported values of $\chi^2_{\rm min}$,  
the first case, assuming the LMA solution, is favoured compared to the second case, assuming the SMA solution, and thus values of $\sin^{2}\alpha \lesssim 10^{-4}$ are 
disfavoured approximately at the $90\%\,c.l.$. In the next year the SNO
experiment should be able to constraint more significantly pure active to 
sterile neutrino oscillations in solar neutrinos and in particular the case 
when the SMA solution is considered, 
unless evidence for $\nu_e \rightarrow \nu_s$ will be found.

Therefore we arrive to the conclusion that {\em current experiments
favour those four neutrino mixing models, both
of 3+1 and 2+2 type, in which the sterile neutrino flavor is  
brought to a complete, or almost complete, thermalization and
no large electron asymmetry generation is possible in a way that
the final result is always $\Delta N_{\nu}^{\rho}=\Delta N_{\nu}^{\rm tot}\simeq 1$.  
Therefore, from the upper limit $\Delta N_{\nu}^{\rm tot}\leq 0.3$, 
current cosmological observations disfavour 4 neutrino mixing models}.
There are however some possibilities for which the cosmological bound
could be evaded.
\begin{itemize}
\item systematic uncertainties or statistical errors have been underestimated in
      $Y_p$ and/or $\eta^{CMB}$ measurements. In the case of higher $Y_p$ and/or lower
      $\eta^{CMB}$ then $\Delta N_{\nu}^{\rm tot}=1$ could be possible. 
      For example one total extra neutrino species would be allowed at $3\,\sigma$ 
      in the case of underestimated systematic uncertainties if: 
      $Y_p=0.254\pm 0.002$ and $\eta^{CMB}$ unchanged as in the (\ref{eta})
      or if $\eta^{CMB}=3.5^{+1.1}_{-0.8}$ and high values $Y_p^{\rm exp}$ are used; 
      in the case of underestimated statistical errors if 
      $Y_p^{\rm exp}=0.244\pm 0.006$ and $\eta^{CMB}$ as in the (\ref{eta})       
      or if $\eta^{CMB}=6.0 \pm 1.5$ and high values of $Y_p^{\rm exp}$ are used  
      \footnote
       {These are qualitative estimations because we are 
      calculating the $99\%$ c.l. just multiplying by a factor 3 the
      error at $68\% c.l.$, as for a gaussian distribution. This is a 
      a very rough assumption when $\delta \eta/\eta$ is not $\ll 1$ and a more
      elaborate statistical procedure should be used.}.
\item We assumed small initial neutrino asymmetries. If some unknown mechanism
      created large neutrino asymmetries ($\sim 10^{-5}-10^{-4}$) 
      above $T\sim 15\,{\rm MeV}$ (the characteristic temperature for                                                                                               oscillations with $|\delta m^2|=\delta m^2_{LSND}\sim 1\,{\rm eV}^2$),  
      then the sterile neutrino production would be completely suppressed \cite{fv}.
      In this case the constraints that we obtained in section 2 should be applied to
      the values of $\Delta N_{\nu}^{\rho}$ and  $\Delta N_{\nu}^{f_{\nu_e}}$ 
      associated to large neutrino asymmetries (see Eq. (\ref{Nxi}) ).
\item We neglected completely the presence of phases in the four neutrino mixing matrix
      since we used the analogy with two neutrino mixing to calculate the cosmological output.                          The role of phases in cosmology has never been studied.
      A possibility could be that, when phases are taken into account, then the
      active-sterile neutrino mixing, even with large angles $\sin^2 2\alpha \gtrsim 10^{-4}$,
      can generate a large neutrino asymmetry that suppresses the sterile neutrino                         production, or in the case of an electron neutrino asymmetry, yields a negative
      $\Delta N_{\nu}^{f_{\nu_e}}$. This possibility should be
      verified in a full 4 neutrino mixing kinetic theory.
\end{itemize}

\section{Five neutrino mixing}

If one assumes the existence of a mixing with 
a second light sterile neutrino flavor $\nu_{s'}$
then it is possible to evade the cosmological bound
if the mixing generates a large neutrino asymmetry able
to suppress the production of the first sterile neutrino 
and in the case of an electron neutrino asymmetry also 
to yield a large negative $\Delta N_{\nu}^{f_{\nu_e}}$.
 This new mass eigenstate should be 
added to the four neutrino mixing solutions that 
explain the experiments and that we described in the previous section.
For convenience let us refer to the 
first neutrino flavor as the LSND neutrino.
It is necessary to distinguish between models in which the LSND
neutrino production is resonant and models in which it is not resonant. 

In the {\em non resonant case},  even though the asymmetry can 
start to be generated by the mixing with the $s'$-neutrino, 
afterwards it gets destroyed by the mixing with the 
LSND neutrino \cite{fv,dll}. Thus the addition of a second sterile
neutrino flavour to the A class `3+1' type models and to the
`2+2' models that are close to the limit cases 2 and 3 
cannot help to evade the cosmological bound.

In the {\em resonant case} the generation of a neutrino
asymmetry from the mixing with $\nu_{s'}$ is not
obstacled by the mixing with the LSND neutrino. Thus any 
$\alpha$-neutrino asymmetry generation  has the effect to suppress the sterile neutrino 
production. However there cannot be a full suppression, because necessarily 
$|\delta m^2_{\alpha s'}|\simeq |\delta m^2_{\alpha s}|$ and the asymmetry generation
starts when already about half of the sterile neutrino production occurred and
the final result is $\Delta N_{\nu}^{\rho}\simeq 0.5$.
This is the only effect in the case of generation of a muon or tauon neutrino asymmetry 
 and thus $\Delta N_{\nu}^{\rm tot}=\Delta N_{\nu}^{\rho}\simeq 0.5$.
This means that adding a second sterile neutrino flavor to the  `2+2' models 
`close' to the limit case 1 (those for $\sin^2\alpha\simeq 0.05$) 
improves the agreement with cosmology but still not within $3\sigma$.
In the case of an electron asymmetry generation one can have
also a negative contribution from $\Delta N_{\nu}^{f_{\nu_e}}$ and the
cosmological bound can be fully evaded. 
This means that the addition of a second sterile neutrino flavor makes possible
to evade the cosmological bound only when it is added to four neutrino spectra
of type `3+1' class B or `2+2' models close to the limit case 4
($\sin^2\alpha \simeq 0.80$), in which the LSND neutrino is present mainly 
in the light pair of mass eigenstates and an electron neutrino asymmetry
can be generated. This possibility to evade the bound 
can be tested both with future $\beta\beta_{0\nu}$ decay experiments but also
with future cosmological CMB observations that should find
$(\Delta N_{\nu}^{\rho})^{CMB}\simeq 0.5$. 
Moreover one should also have  $\eta^{CMB}>\eta^{SBBN}_{D/H}$, 
but considering the current error on $D/H$ measurement,
this possibility can be distinguished only if future CMB observations will give
$\eta^{CMB}\gtrsim 7.7$ (at $3\sigma$). 

\section{Discussion and conclusions}

We described an analytical and graphical procedure to search 
for non standard effects  from nuclear abundances and CMB obervations. 
The recent measurement of the baryon content from CMB anisotropies 
improves remarkably the cosmological information on new physics. 
The present observations do not show evidences of the presence of non 
standard effects and constraints can be conveniently displayed in 
the $\Delta N_{\nu}^{\rm tot}-\Delta N_{\nu}^{\rho}$ plot.   

However future measurements of $\eta$ and $\Delta N_{\nu}^{\rho}$
from CMB, together with the current 
measurements of primordial Helium-4 and Deuterium nuclear abundances,
could provide some interesting signatures. Here is a summary list of the 
possible signatures as we found in the second section.

\begin{enumerate}
\item If $\eta^{CMB}\gtrsim 7$ then
      $\Delta N_{\nu}^{\rm tot}<0$ and a negative $\Delta N_{\nu}^{f_{\nu_e}}$
      can be invoked.
\item  If $\eta^{CMB}\gtrsim 7.7$ then also $(\Delta N_{\nu}^{\rho})^{BBN}>0$.
\item  If $(\Delta N_{\nu}^{\rho})^{CMB}>0$ then $(\Delta N_{\nu}^{\rho})^{BBN}>0$
       if one can exclude massive neutrino decays or other exotic effects 
       intervening between the BBN and recombination epochs. 
\item If ($\Delta N_{\nu}^{\rho})^{BBN}\gtrsim 0.3$ then, from the bound 
      $\Delta N_{\nu}^{\rm tot}< 0.3$, one can conclude that 
      $\Delta N_{\nu}^{f_{\nu_e}}<0$.
\item If the point 2 is verified but 
      $(\Delta N_{\nu}^{\rho})^{CMB}<(\Delta N_{\nu}^{\rho})^{BBN}$, then
      this can be interpreted as a signature of inhomogenous 
      $\Delta N_{\nu}^{f_{\nu_e}}$. This should be confirmed by  
	inhomogeneities in $(D/H)$ measurements that should be anyway
      searched independently on CMB observations.
\end{enumerate} 

We have applied these cosmological tools to the search of non standard effects
from neutrino mixing. In the case of three ordinary neutrino mixing 
it seems not possible to find relevant cosmological effects. 
When a mixing with new light sterile neutrino flavors
is considered, like the LSND experiment seems to suggest, then the early Universe
becomes a powerful probe. We have shown how the SNO and SK experiments
favour those four neutrino mixing models for which
the sterile neutrino flavor is brought into thermal equilibrium or very close to it,
implying that $\Delta N_{\nu}^{\rho}\simeq 1$. At the same time 
a mechanism of electron neutrino asymmetry generation cannot be invoked
to have a negative $\Delta N_{\nu}^{f_{\nu_e}}$ and thus in the end
$\Delta N_{\nu}^{\rm tot}\simeq 1$. The cosmological
bound $\Delta N_{\nu}^{\rm tot}<0.3$ is already quite conservative and 
future cosmological observations will be compatible with
$\Delta N_{\nu}^{\rm tot}\simeq 1$ only if  they will measure 
a value for $\eta^{CMB}$ that should be approximately half than
the value measured by current observations or, alternatively,
a value of $Y_p$ that should be about 0.01 higher. This of course
would mean that present observations are affected by large
systematic uncertainties or that statistical errors have been largely 
underestimated. If one excludes such a possibility then a 
way out could be the presence of large initial neutrino 
asymmetries suppressing the sterile neutrino production. In this case the cosmological 
information can still be used to constraint the values of the neutrino asymmetries. 
Such a conclusion would have a quite remarkable impact on 
baryo-leptogenesis models.
Another possibility is that phases in the neutrino mixing matrix could play
an important role in the derivation of cosmological output and thus should be taken into
account. Another intriguing possibility is to assume the existence of more than one sterile neutrino flavor. The new sterile neutrino flavor
should be mixed with the electron neutrino flavor with the proper mixing parameters 
such that a relevant electron neutrino asymmetry is generated and both halves
the sterile neutrino production and yields a negative $\Delta N_{\nu}^{f_{\nu_e}}$. 
This is possible only if the electron neutrino flavor is mainly
present in the heavy mass eigenstates with $m_{i}\sim 1\,{\rm eV}$. Therefore
this scenario will be testable  in future $\beta\beta_{0\nu}$ decay experiments
and with the cosmological tools that we described. 

This investigation thus shows that light sterile neutrinos in cosmology are
now more constrained than before, because the possibility of
a neutrino asymmetry generation in four neutrino mixing models is disfavoured 
within the statistical significance of the results from
the SNO \cite{SNO} and the SK \cite{SK} experiments. The  
result is that the sterile neutrino flavor, required by the LSND experiment,
gets fully thermalized. Therefore the confirmation of the existence of light 
sterile neutrino flavors in next neutrino mixing experiments would be of
great relevance for cosmology. Such a confirmation
should come in next years by many planned experiments. In particular the
MiniBooNE  experiment should confirm or disprove the evidence of neutrino 
oscillations in the LSND experiment, 
while many other different experiments will be able to exclude exotic solutions 
to explain solar and atmospheric neutrino data.

\vskip 0.5cm
\noindent
{\bf Acknowledgements}

The author is an Alexander von Humboldt research fellow. He wishes to thank
L. Mersini for many valuable comments and discussions. He also
thanks R. Foot and R.R. Volkas for encouraging comments, M.C. 
Gonzalez-Garcia for a useful discussion during the EPS HEP 2001 conference in Budapest, 
M. Tegmark and X. Wang for explanations on CMB data analysis, 
Q. Shafi and A. Ringwald for nice discussions. He is grateful to M. Lusignoli
for having drawn to his attention important points on
the statistical significance of current four neutrino mixing  data 
analysis in excluding pure active to sterile neutrino oscillations.

\newpage
{\bf\large Figure Captions}
\vskip 0.5cm
\noindent
{\bf Figure 1.} Constraints on non standard BBN models from 
measurements of $\eta^{(CMB)}$,  $Y_p$ and $(D/H)$. The solid vertical 
lines are the constraints (\ref{Ntoth}) and (\ref{Ntotl}) with the thick
ones indicating the joint range coming from low+high $Y_p$ values.
The horizontal solid lines is the constraint (\ref{Nrho}).
The dark gray region is the allowed one by current observations.
The dashed lines, contours of the gray region, are the constraints obtained 
neglecting the error on $\eta^{CMB}$ in the BBN predictions
and assuming a value $\eta^{(CMB)}=\eta^{SBBN}=5.6$,
corresponding to $(D/H)=3.0\times 10^{-5}$ in SBBN. 
The light gray region is the allowed region if one assumes 
$\eta^{CMB}=7$ and low+high $Y_p^{\rm exp}$ range of values.
The dotted horizontal lines are the realistic constraints 
that will be obtained on $\Delta N_{\nu}^{\rho}$ from 
future CMB anisotropy observations. The BBN from 
Standard Model of Particle Physics lies well 
within the circle centered around the origin.

\noindent
{\bf Figure 2.} Accessible region for $\nu_{\alpha}\leftrightarrow\nu_s$
in the plane $\Delta N_{\nu}^{\rm tot}-\Delta N_{\nu}^{\rho}$. 
The thick segment along the line $\Delta N_{\nu}^{\rm tot}=\Delta N_{\nu}^{\rho}$
corresponds to the case $\alpha=\mu,\tau$, while the striped regions are for
the case $\alpha=e$. The solid striped region is accessible in the case
of $\nu_e,\bar{\nu}_e$ number density depletion below the neutrino chemical decoupling 
or in the case of {\em negative} electron asymmetry generation ($\Delta N_{\nu}^{f_{\nu_e}}\geq 0$) plus sterile neutrino production ($\Delta N_{\nu}^{\rho}\geq 0$). 
The dashed striped region is accessible when a large {\em positive} electron neutrino 
asymmetry is generated. The thick dashed segment for
$\Delta N_{\nu}^{\rho}=0$ corresponds to the region of mixing parameters 
for which the sterile neutrino production in the collision dominated regime is 
negligible. The possibility to have both an asymmetry generation and a sterile neutrino
production ($\Delta N_{\nu}^{f_{\nu_e}}\neq 0$, $\Delta N_{\nu}^{\rho}> 0$)
has not been studied in detail and there are only 
particular numerical examples. The dotted line 
is a simple interpolation between the 
two extreme cases $\Delta N_{\nu}^{\rho}=0$ and $\Delta N_{\nu}^{\rho}=1$
that provides a reasonable approximation.

\noindent
{\bf Figure 3.} Approximate accessible regions in the
$\Delta N_{\nu}^{\rm tot}-\Delta N_{\nu}^{\rho}$ plot,
in the case of four neutrino mixing models.
The solid lines are for A class `2+2' models.
The region between the dashed lines constraints B class `2+2' models. 
The dotted segment is  for the A class `3+1' model. 
The small circle is for the B class `3+1' model and for all `2+2' models
when the information from SNO and SK is used and it can be seen that
it lies well outside the allowed cosmological region (in gray).

\newpage
\psfig{file=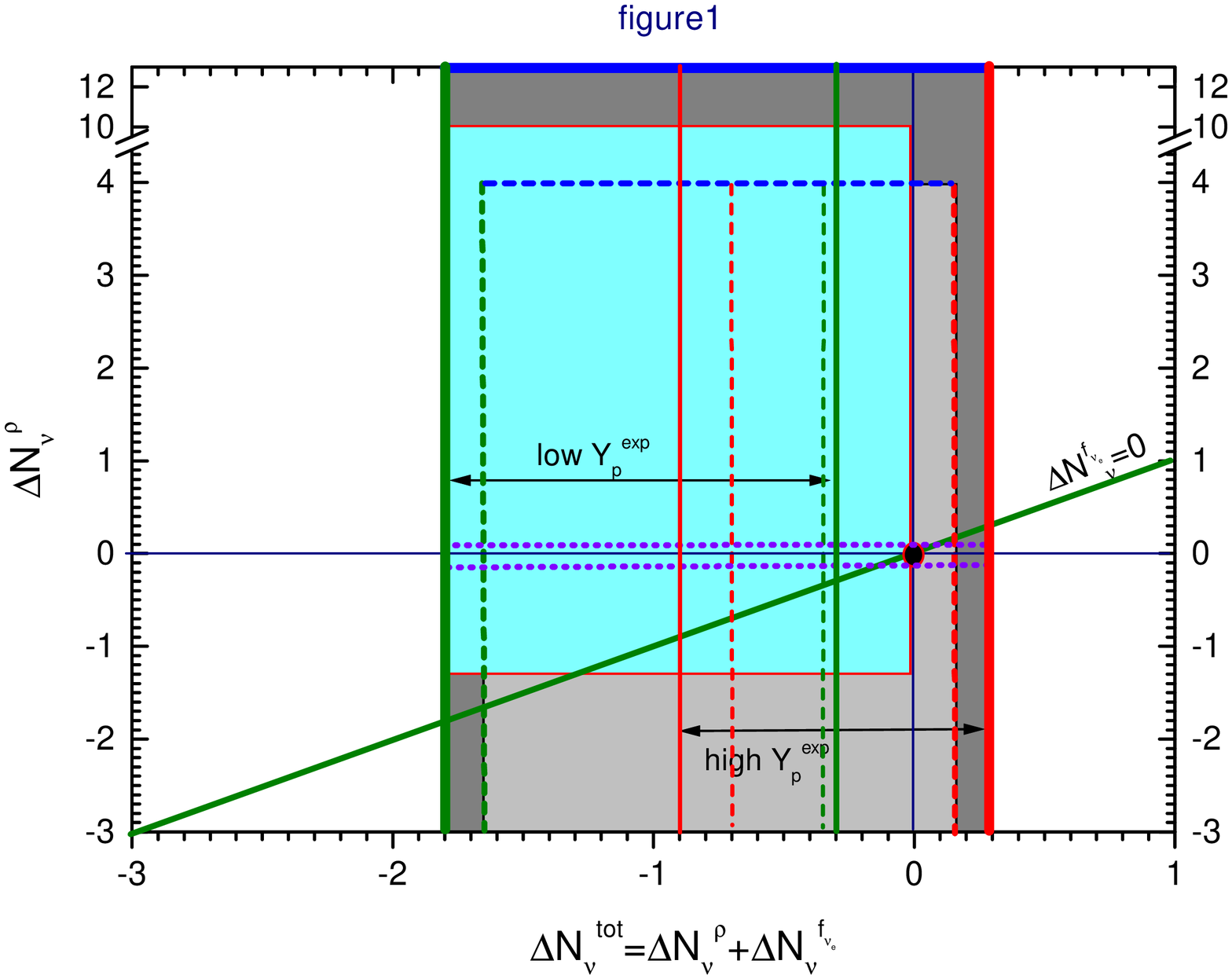,width=15cm}
\newpage
\psfig{file=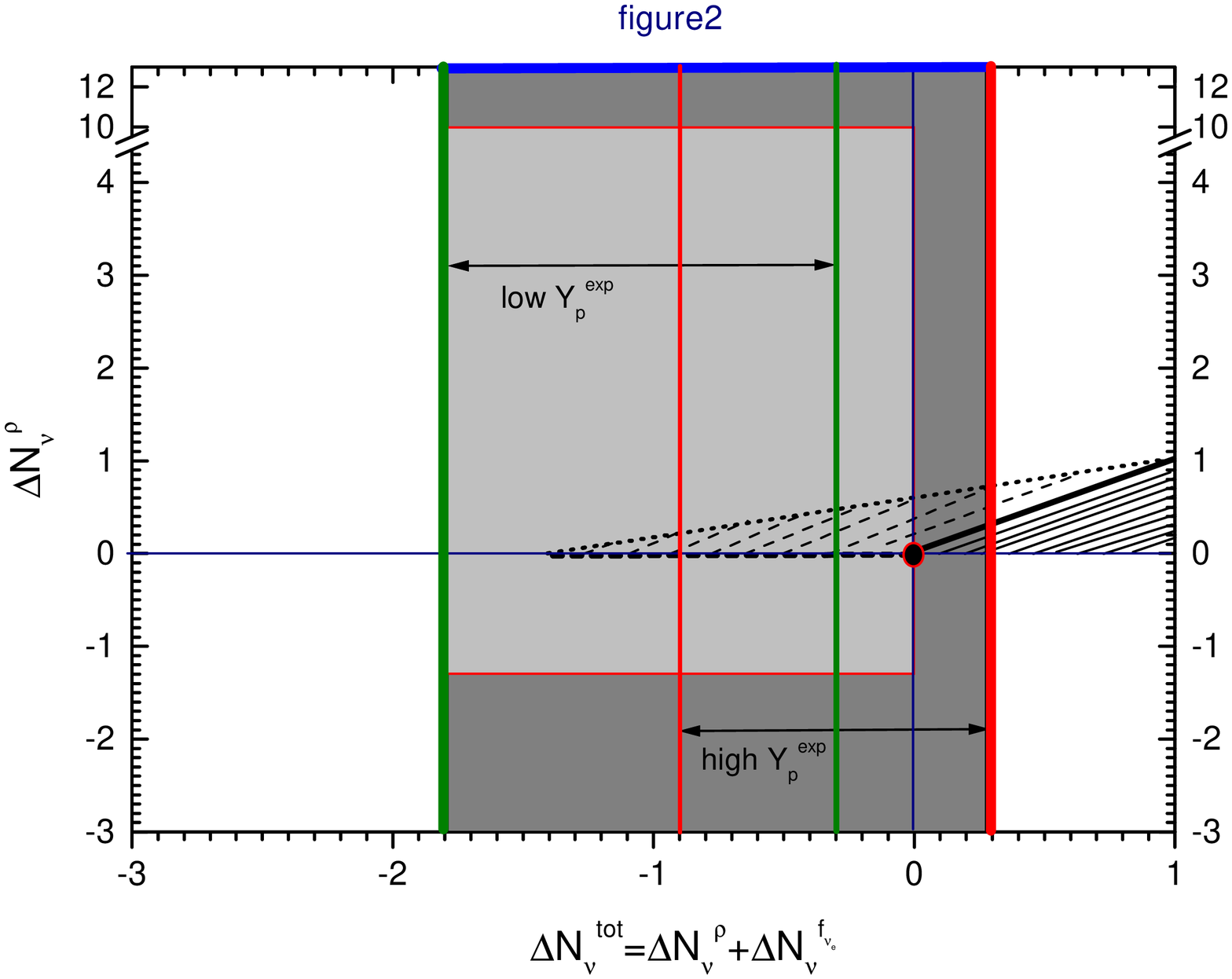,width=15cm}
\newpage
\psfig{file=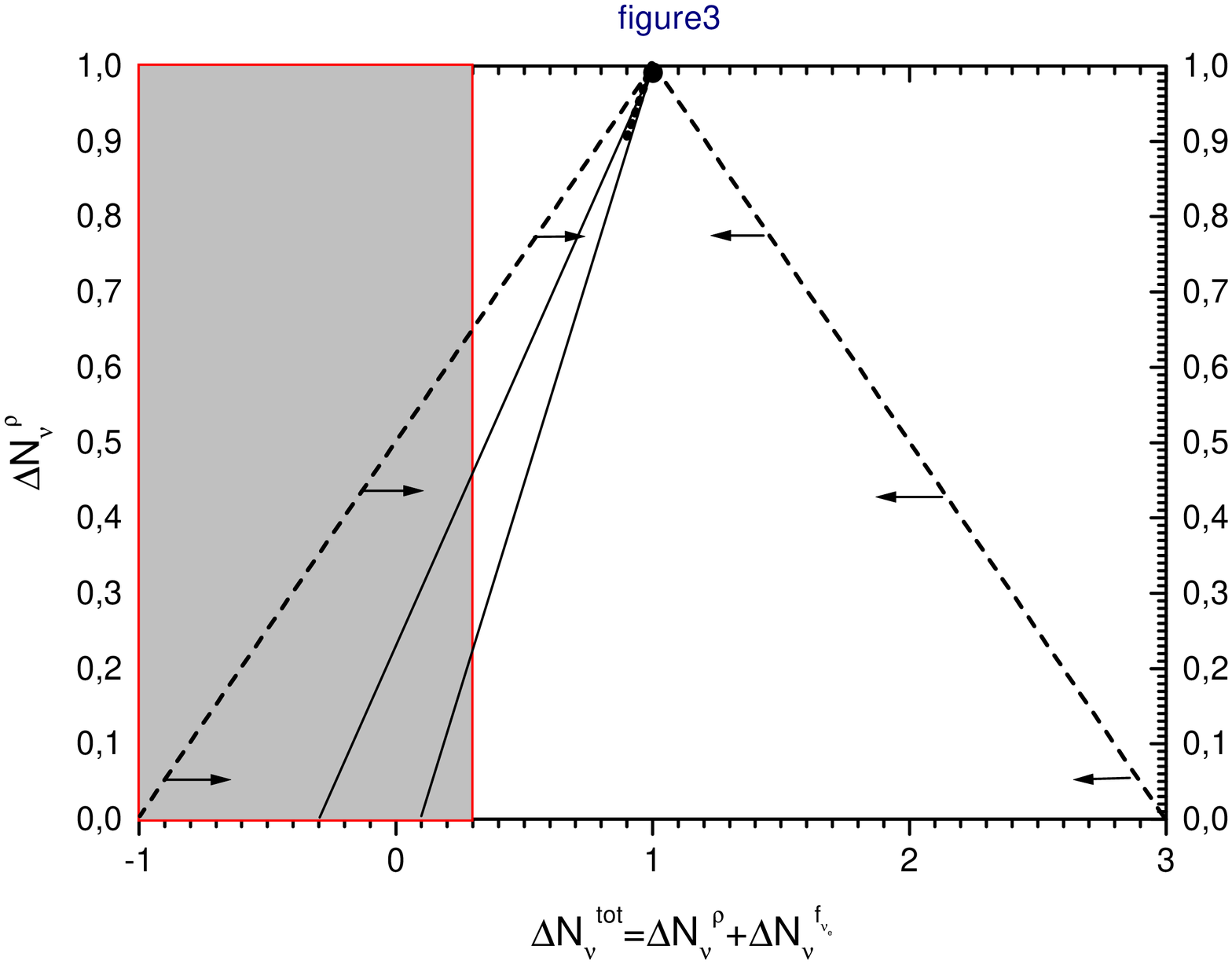,width=15cm}


\begin{thebibliography}{10}

\bibitem{atm}
  Y. Fukuda \textit{et al.}, Super-Kamiokande Coll.,
  Phys. Rev. Lett. \textbf{81} (1998) 1562; Phys.\ Rev.\ Lett.\ {\bf
    82}, 2644 (1999); Y. Fukuda \textit{et al.},
  Kamiokande Coll., Phys. Lett. B \textbf{335} (1994) 237;
  R. Becker-Szendy \textit{et al.},
  IMB Coll., Nucl. Phys. B (Proc. Suppl.) \textbf{38} (1995) 331;
  W.W.M. Allison \textit{et al.}, Soudan Coll.,
  Phys. Lett. B \textbf{449} (1999) 137;
  M. Ambrosio \textit{et al.},
  MACRO Coll., Phys. Lett. B \textbf{434} (1998) 451.
  C. McGrew in {\it Neutrino Telescopes 2001}, Venice, Italy, March 2001,
  to appear; T. Toshito in {\it Moriond 2001}, Les Arcs, France, March 2001,
  to appear. For a review see T.~Kajita and Y.~Totsuka,
  Rev.\ Mod.\ Phys.\  {\bf 73} (2001) 85.

\bibitem{solar}
  B.T. Cleveland \textit{et al.}, Astrophys. J. \textbf{496} (1998) 505;
  K.S. Hirata \textit{et al.},
  Kamiokande Coll., Phys. Rev. Lett. \textbf{77} (1996) 1683;
  W. Hampel \textit{et al.},
  GALLEX Coll., Phys. Lett. B \textbf{447} (1999) 127;
  D.N. Abdurashitov \textit{et al.},
  SAGE Coll., Phys. Rev. Lett. \textbf{83} (1999) 4686;
  Y. Fukuda \textit{et al.},
  Super-Kamiokande Coll., Phys. Rev. Lett. \textbf{81} (1998) 1158.
  
\bibitem{SNO}
Q.~R.~Ahmad {\it et al.}  [SNO Collaboration],
nucl-ex/0106015.



\bibitem{noexotic}
  P.~Lipari and M.~Lusignoli,
  Phys.\ Rev.\ D {\bf 60} (1999) 013003 [hep-ph/9901350].
  G.~L.~Fogli, E.~Lisi, A.~Marrone and G.~Scioscia, Phys.\ Rev.\ D {\bf 60} (1999)
  053006 [hep-ph/9904248]; 
N. Fornengo, M. Maltoni, R. Tom{\`a}s Bayo and J. W. F. Valle, hep-ph/0108043.

\bibitem{LSND}
A.~Aguilar {\it et al.}  [LSND Collaboration],
hep-ex/0104049.

\bibitem{Shvartsman}
V.~F.~Shvartsman,
Pisma Zh.\ Eksp.\ Teor.\ Fiz.\  {\bf 9} (1969) 315
[JETP Lett.\  {\bf 9} (1969) 184].

\bibitem{Schramm}
G.~Steigman, D.~N.~Schramm and J.~R.~Gunn,
Phys.\ Lett.\ B {\bf 66} (1977) 202.

\bibitem{CMB}
P.~de Bernardis {\it et al.}, Nature {\bf 404} (2000) 955 [astro-ph/0004404];
S.~Hanany {\it et al.},
Astrophys.\ J.\  {\bf 545} (2000) L5
[astro-ph/0005123];
N.~W.~Halverson {\it et al.}, astro-ph/0104489.

\bibitem{CMB2}
A.~E.~Lange {\it et al.},
Phys.\ Rev.\ D {\bf 63} (2001) 042001
[astro-ph/0005004];
A.~Balbi {\it et al.},
Astrophys.\ J.\  {\bf 545} (2000) L1 [astro-ph/0005124].

\bibitem{BOOM}
C.~B.~Netterfield {\it et al.},
astro-ph/0104460.

\bibitem{DASI}
C.~Pryke {\it et al.}, astro-ph/0104490.

\bibitem{MAX}
R.~Stompor {\it et al.}, astro-ph/0105062.

\bibitem{MAP}
http://map.gsfc.nasa.gov/m\_mm/ms\_status.html

\bibitem{PLANCK}
http://astro.estec.esa.nl/Planck/

\bibitem{WTZ}
X.~Wang, M.~Tegmark and M.~Zaldarriaga, astro-ph/0105091. 



\bibitem{private}
M.~Tegmark, private communication.



\bibitem{Wagoner}
R.~V.~Wagoner, W.~A.~Fowler and F.~Hoyle,
Astrophys.\ J.\  {\bf 148} (1967) 3;
M.~S.~Smith, L.~H.~Kawano and R.~A.~Malaney,
Astrophys.\ J.\ Supp.\  {\bf 85} (1993) 219.


\bibitem{Izotov} 
Y.I. Izotov and  T.X. Thuan, Ap.J. {\bf 500}, 188 (1998).

\bibitem{Olive}
K.~A.~Olive and G.~Steigman,
Astrophys.\ J.\ Supp.\  {\bf 97} (1995) 49
[astro-ph/9405022].

\bibitem{O'Meara}
J.~M.~O'Meara, D.~Tytler, D.~Kirkman, N.~Suzuki, J.~X.~Prochaska, D.~Lubin and A.~M.~Wolfe,
astro-ph/0011179.

\bibitem{Lopez}
R.~E.~Lopez and M.~S.~Turner,
Phys.\ Rev.\ D {\bf 59} (1999) 103502
[astro-ph/9807279].

\bibitem{Walker}
T.~P.~Walker, G.~Steigman, D.~N.~Schramm, K.~A.~Olive and H.~Kang,
Astrophys.\ J.\  {\bf 376} (1991) 51.

\bibitem{Sarkar96}
S.~Sarkar,
Rept.\ Prog.\ Phys.\  {\bf 59} (1996) 1493
[hep-ph/9602260].

\bibitem{Dolgov}
A.~D.~Dolgov, S.~H.~Hansen and D.~V.~Semikoz,
Nucl.\ Phys.\ B {\bf 503} (1997) 426
[hep-ph/9703315].

\bibitem{Lisi}
E.~Lisi, S.~Sarkar and F.~L.~Villante,
Phys.\ Rev.\ D {\bf 59} (1999) 123520
[hep-ph/9901404].

\bibitem{Andreas}
Z.~Fodor, S.~D.~Katz and A.~Ringwald,
hep-ph/0105064.

\bibitem{Hannu}
H.~Kurki-Suonio and E.~Sihvola,
Phys.\ Rev.\ D {\bf 63} (2001) 083508
[astro-ph/0011544].

\bibitem{Bond}
G.~Jungman, M.~Kamionkowski, A.~Kosowsky and D.~N.~Spergel,
Phys.\ Rev.\ D {\bf 54} (1996) 1332 [astro-ph/9512139];
J.~R.~Bond, G.~Efstathiou and M.~Tegmark, 
Mon.Not.R.Astron.Soc. {\bf 291}, 33 (1997).

\bibitem{Hata}
N.~Hata, R.~J.~Scherrer, G.~Steigman, D.~Thomas, T.~P.~Walker, S.~Bludman and P.~Langacker,
Phys.\ Rev.\ Lett.\  {\bf 75} (1995) 3977
[hep-ph/9505319].

\bibitem{Hannestad}
S.~Hannestad,
astro-ph/0105220.

\bibitem{Lopez2}
R.~E.~Lopez, S.~Dodelson, A.~Heckler and M.~S.~Turner,
Phys.\ Rev.\ Lett.\  {\bf 82} (1999) 3952
[astro-ph/9803095].

\bibitem{Hannestad2}
S.~Hannestad,
Phys.\ Rev.\ Lett.\  {\bf 85} (2000) 4203
[astro-ph/0005018].

\bibitem{Pagel}
A.~D.~Dolgov and B.~E.~Pagel,
New Astron.\  {\bf 4} (1999) 223
[astro-ph/9711202].

\bibitem{Reeves}
H.~Reeves, Phys.\ Rev.\ D {\bf 6} (1972) 3363;
A.~Yahil and G.~Beaudet, Astrophys.\ J.\  {\bf 206} (1976) 26;
G. Beaudet and P. Goret, Astron. \& Astrophys. {\bf 49}, 415 (1976);
K.~A.~Olive, D.~N.~Schramm, D.~Thomas and T.~P.~Walker,
Phys.\ Lett.\ B {\bf 265} (1991) 239;
H.~Kang and G.~Steigman,
Nucl.\ Phys.\ B {\bf 372} (1992) 494.

\bibitem{Kneller}
J.~P.~Kneller, R.~J.~Scherrer, G.~Steigman and T.~P.~Walker,
astro-ph/0101386;
J.~Lesgourgues and A.~R.~Liddle,
astro-ph/0105361.
S.~H.~Hansen, G.~Mangano, A.~Melchiorri, G.~Miele and O.~Pisanti,
astro-ph/0105385.

\bibitem{CHOOZ}
M.~Apollonio {\it et al.}  [CHOOZ Collaboration],
Phys.\ Lett.\ B {\bf 466} (1999) 415
[hep-ex/9907037].

\bibitem{Fogli}
G.~L.~Fogli, E.~Lisi, D.~Montanino and A.~Palazzo, hep-ph/0106247;
A.~Bandyopadhyay, S.~Choubey, S.~Goswami and K.~Kar,
arXiv:hep-ph/0106264.

\bibitem{Bahcall}
J.~N.~Bahcall, M.~C.~Gonzalez-Garcia and C.~Pena-Garay,
JHEP {\bf 0108} (2001) 014
[hep-ph/0106258].

\bibitem{bimaximal}
F.~Vissani, hep-ph/9708483;
V.~Barger, S.~Pakvasa, T.~J.~Weiler and K.~Whisnant,
Phys.\ Lett.\ B {\bf 437} (1998) 107
[hep-ph/9806387].
A.~J.~Baltz, A.~S.~Goldhaber and M.~Goldhaber,
Phys.\ Rev.\ Lett.\  {\bf 81} (1998) 5730
[hep-ph/9806540];
G.~Altarelli and F.~Feruglio, Phys.\ Lett.\ B {\bf 439} (1998) 112
[hep-ph/9807353].

\bibitem{Langacker}
P.~Langacker, S.~T.~Petcov, G.~Steigman and S.~Toshev,
Nucl.\ Phys.\ B {\bf 282} (1987) 589.




\bibitem{Peltoniemi}
J.~T.~Peltoniemi, D.~Tommasini and J.~W.~Valle,
Phys.\ Lett.\ B {\bf 298} (1993) 383;
J.~T.~Peltoniemi and J.~W.~Valle,
Nucl.\ Phys.\ B {\bf 406} (1993) 409
[hep-ph/9302316];
D.~O.~Caldwell and R.~N.~Mohapatra,
Phys.\ Rev.\ D {\bf 48} (1993) 3259.

\bibitem{PDG}
D.~E.~Groom {\it et al.}  [Particle Data Group Collaboration],
Eur.\ Phys.\ J.\ C {\bf 15} (2000) 1.

\bibitem{fv}
R.~Foot and R.~R.~Volkas,
Phys.\ Rev.\ Lett.\ {\bf 75} (1995) 4350
[hep-ph/9508275].

\bibitem{dll}
P.~Di Bari, P.~Lipari and M.~Lusignoli,
Int.\ J.\ Mod.\ Phys.\ A {\bf 15} (2000) 2289 [hep-ph/9907548].

\bibitem{Cline}
J.~M.~Cline, Phys.\ Rev.\ Lett.\  {\bf 68} (1992) 3137.

\bibitem{ftv}
R.~Foot, M.~J.~Thomson and R.~R.~Volkas,
Phys.\ Rev.\ {\bf D 53} (1996) 5349
[hep-ph/9509327];
R.~Foot and R.~R.~Volkas,
Phys.\ Rev.\ {\bf D 55}, 5147 (1997)
[hep-ph/9610229].

\bibitem{comment}
P.~Di Bari, R.~Foot, R.~R.~Volkas and Y.~Y.~Wong,
hep-ph/0008245.

\bibitem{ekt}
K.~Enqvist, K.~Kainulainen and M.~Thomson,
Nucl.\ Phys.\ B {\bf 373} (1992) 498.

\bibitem{Dolgov4}
A.~D.~Dolgov,
Phys.\ Lett.\ B {\bf 506} (2001) 7
[hep-ph/0006103].

\bibitem{fv2}
R.~Foot and R.~R.~Volkas,
Phys.\ Rev.\ {\bf D 56}, 6653 (1997)
[hep-ph/9706242].

\bibitem{ropa3}
P.~Di Bari and R.~Foot,
hep-ph/0103192.

\bibitem{MS}
S.~P.~Mikheev and A.~Y.~Smirnov,
Nuovo Cim.\ C {\bf 9} (1986) 17;
Sov.\ J.\ Nucl.\ Phys.\  {\bf 42} (1985) 913
[Yad.\ Fiz.\  {\bf 42} (1985) 1441].

\bibitem{Robert}
R.~Foot, Phys.\ Rev.\ D {\bf 61} (2000) 023516
[hep-ph/9906311].

\bibitem{Kirillova}
D.~P.~Kirilova and M.~V.~Chizhov, Nucl.\ Phys.\ B {\bf 591} (2000) 457
[hep-ph/9909408].

\bibitem{Kirillova2}
D.~P.~Kirilova, astro-ph/0109105.

\bibitem{constraints}
K.~Kainulainen,
Phys.\ Lett.\ B {\bf 244} (1990) 191;
R.~Barbieri and A.~Dolgov,
Phys.\ Lett.\ B {\bf 237} (1990) 440.
X.~Shi, D.~N.~Schramm and B.~D.~Fields,
Phys.\ Rev.\ D {\bf 48} (1993) 2563
[astro-ph/9307027].
D.~P.~Kirilova and M.~V.~Chizhov,
Phys.\ Lett.\ B {\bf 393} (1997) 375
[hep-ph/9608270];

\bibitem{Shi}
X.~Shi, Phys.\ Rev.\ D {\bf 54} (1996) 2753
[arXiv:astro-ph/9602135].

\bibitem{ropa2}
P.~Di Bari and R.~Foot,
Phys.\ Rev.\ D {\bf 63} (2001) 043008 [hep-ph/0008258].

\bibitem{BoomMax}
A.H.~Jaffe et.al., Phys. Rev. Lett. {\bf 86} (2001) 3475.

\bibitem{domains}
P.~Di Bari,
Phys.\ Lett.\ B {\bf 482} (2000) 150
[hep-ph/9911214].

\bibitem{Grasso}
A.~D.~Dolgov and D.~Grasso, astro-ph/0106154.

\bibitem{Okada}
N.~Okada and O.~Yasuda, Int.\ J.\ Mod.\ Phys.\ A {\bf 12} (1997) 3669
[hep-ph/9606411].

\bibitem{Bilenky}
S.~M.~Bilenkii, C.~Giunti and W.~Grimus,
Eur.\ Phys.\ J.\ C {\bf 1} (1998) 247
[hep-ph/9607372];

\bibitem{Barger0}
V.~Barger, S.~Pakvasa, T.~J.~Weiler and K.~Whisnant,
Phys.\ Rev.\ D {\bf 58} (1998) 093016
[hep-ph/9806328].

\bibitem{others}
S.~M.~Bilenkii, C.~Giunti, W.~Grimus and T.~Schwetz,
Phys.\ Rev.\ D {\bf 60} (1999) 073007
[hep-ph/9903454]; 
C.~Giunti and M.~Laveder,
JHEP {\bf 0102} (2001) 001
[hep-ph/0010009];
O.~L.~Peres and A.~Y.~Smirnov,
Nucl.\ Phys.\ B {\bf 599} (2001) 3
[hep-ph/0011054];
W.~Grimus and T.~Schwetz,
Eur.\ Phys.\ J.\ C {\bf 20} (2001) 1
[hep-ph/0102252].

\bibitem{Barger1}
V.~Barger, B.~Kayser, J.~Learned, T.~Weiler and K.~Whisnant,
Phys.\ Lett.\ B {\bf 489} (2000) 345
[hep-ph/0008019];


\bibitem{Valle}
M.~Maltoni, T.~Schwetz and J.~W.~Valle,
hep-ph/0107150;

\bibitem{Gonzalez}
M.~C.~Gonzalez-Garcia, M.~Maltoni and C.~Pena-Garay,
hep-ph/0108073.
For earlier works (pre-SNO results \cite{SNO}) see for example:
C.~Giunti, M.~C.~Gonzalez-Garcia and C.~Pena-Garay,
Phys.\ Rev.\ D {\bf 62} (2000) 013005
[hep-ph/0001101];
O.~Yasuda,
hep-ph/0006319;
M.~Hirsch and J.~W.~Valle,
Phys.\ Lett.\ B {\bf 495} (2000) 121
[hep-ph/0009066].


\bibitem{SK}
S.~Fukuda {\it et al.}  [Super-Kamiokande Collaboration],
Phys.\ Rev.\ Lett.\  {\bf 85} (2000) 3999
[hep-ex/0009001].


\bibitem{Rob}
R.~Foot, Phys.\ Lett.\ B {\bf 496} (2000) 169
[hep-ph/0007065].



\bibitem{Barger}
V.~Barger, D.~Marfatia and K.~Whisnant,
hep-ph/0106207.

\bibitem{Klapdor}
H.~V.~Klapdor-Kleingrothaus, H.~Pas and A.~Y.~Smirnov,
Phys.\ Rev.\ D {\bf 63} (2001) 073005
[hep-ph/0003219];
S.~M.~Bilenky, S.~Pascoli and S.~T.~Petcov,
hep-ph/0104218.

\bibitem{Bell}
N.~F.~Bell, R.~Foot and R.~R.~Volkas,
Phys.\ Rev.\ D {\bf 58} (1998) 105010
[hep-ph/9805259].



\end{thebibliography}
\end{document}